\documentclass[a4paper,fleqn,usenatbib]{mnras}

\usepackage{newtxtext,newtxmath}

\usepackage[T1]{fontenc}
\usepackage{ae,aecompl}

\usepackage{graphicx}	
\usepackage{amsmath}	
\usepackage{bm}
\usepackage{subcaption}
\usepackage{tikz}
\usepackage{stfloats}
\usepackage{fontawesome}
\usetikzlibrary{decorations.pathreplacing,arrows,shapes,positioning}  
\usepackage[draft]{minted}
\makeatletter
\def\PYG@reset{\let\PYG@it=\relax \let\PYG@bf=\relax%
    \let\PYG@ul=\relax \let\PYG@tc=\relax%
    \let\PYG@bc=\relax \let\PYG@ff=\relax}
\def\PYG@tok#1{\csname PYG@tok@#1\endcsname}
\def\PYG@toks#1+{\ifx\relax#1\empty\else%
    \PYG@tok{#1}\expandafter\PYG@toks\fi}
\def\PYG@do#1{\PYG@bc{\PYG@tc{\PYG@ul{%
    \PYG@it{\PYG@bf{\PYG@ff{#1}}}}}}}
\def\PYG#1#2{\PYG@reset\PYG@toks#1+\relax+\PYG@do{#2}}

\expandafter\def\csname PYG@tok@w\endcsname{\def\PYG@tc##1{\textcolor[rgb]{0.73,0.73,0.73}{##1}}}
\expandafter\def\csname PYG@tok@c\endcsname{\let\PYG@it=\textit\def\PYG@tc##1{\textcolor[rgb]{0.25,0.50,0.50}{##1}}}
\expandafter\def\csname PYG@tok@cp\endcsname{\def\PYG@tc##1{\textcolor[rgb]{0.74,0.48,0.00}{##1}}}
\expandafter\def\csname PYG@tok@k\endcsname{\let\PYG@bf=\textbf\def\PYG@tc##1{\textcolor[rgb]{0.00,0.50,0.00}{##1}}}
\expandafter\def\csname PYG@tok@kp\endcsname{\def\PYG@tc##1{\textcolor[rgb]{0.00,0.50,0.00}{##1}}}
\expandafter\def\csname PYG@tok@kt\endcsname{\def\PYG@tc##1{\textcolor[rgb]{0.69,0.00,0.25}{##1}}}
\expandafter\def\csname PYG@tok@o\endcsname{\def\PYG@tc##1{\textcolor[rgb]{0.40,0.40,0.40}{##1}}}
\expandafter\def\csname PYG@tok@ow\endcsname{\let\PYG@bf=\textbf\def\PYG@tc##1{\textcolor[rgb]{0.67,0.13,1.00}{##1}}}
\expandafter\def\csname PYG@tok@nb\endcsname{\def\PYG@tc##1{\textcolor[rgb]{0.00,0.50,0.00}{##1}}}
\expandafter\def\csname PYG@tok@nf\endcsname{\def\PYG@tc##1{\textcolor[rgb]{0.00,0.00,1.00}{##1}}}
\expandafter\def\csname PYG@tok@nc\endcsname{\let\PYG@bf=\textbf\def\PYG@tc##1{\textcolor[rgb]{0.00,0.00,1.00}{##1}}}
\expandafter\def\csname PYG@tok@nn\endcsname{\let\PYG@bf=\textbf\def\PYG@tc##1{\textcolor[rgb]{0.00,0.00,1.00}{##1}}}
\expandafter\def\csname PYG@tok@ne\endcsname{\let\PYG@bf=\textbf\def\PYG@tc##1{\textcolor[rgb]{0.82,0.25,0.23}{##1}}}
\expandafter\def\csname PYG@tok@nv\endcsname{\def\PYG@tc##1{\textcolor[rgb]{0.10,0.09,0.49}{##1}}}
\expandafter\def\csname PYG@tok@no\endcsname{\def\PYG@tc##1{\textcolor[rgb]{0.53,0.00,0.00}{##1}}}
\expandafter\def\csname PYG@tok@nl\endcsname{\def\PYG@tc##1{\textcolor[rgb]{0.63,0.63,0.00}{##1}}}
\expandafter\def\csname PYG@tok@ni\endcsname{\let\PYG@bf=\textbf\def\PYG@tc##1{\textcolor[rgb]{0.60,0.60,0.60}{##1}}}
\expandafter\def\csname PYG@tok@na\endcsname{\def\PYG@tc##1{\textcolor[rgb]{0.49,0.56,0.16}{##1}}}
\expandafter\def\csname PYG@tok@nt\endcsname{\let\PYG@bf=\textbf\def\PYG@tc##1{\textcolor[rgb]{0.00,0.50,0.00}{##1}}}
\expandafter\def\csname PYG@tok@nd\endcsname{\def\PYG@tc##1{\textcolor[rgb]{0.67,0.13,1.00}{##1}}}
\expandafter\def\csname PYG@tok@s\endcsname{\def\PYG@tc##1{\textcolor[rgb]{0.73,0.13,0.13}{##1}}}
\expandafter\def\csname PYG@tok@sd\endcsname{\let\PYG@it=\textit\def\PYG@tc##1{\textcolor[rgb]{0.73,0.13,0.13}{##1}}}
\expandafter\def\csname PYG@tok@si\endcsname{\let\PYG@bf=\textbf\def\PYG@tc##1{\textcolor[rgb]{0.73,0.40,0.53}{##1}}}
\expandafter\def\csname PYG@tok@se\endcsname{\let\PYG@bf=\textbf\def\PYG@tc##1{\textcolor[rgb]{0.73,0.40,0.13}{##1}}}
\expandafter\def\csname PYG@tok@sr\endcsname{\def\PYG@tc##1{\textcolor[rgb]{0.73,0.40,0.53}{##1}}}
\expandafter\def\csname PYG@tok@ss\endcsname{\def\PYG@tc##1{\textcolor[rgb]{0.10,0.09,0.49}{##1}}}
\expandafter\def\csname PYG@tok@sx\endcsname{\def\PYG@tc##1{\textcolor[rgb]{0.00,0.50,0.00}{##1}}}
\expandafter\def\csname PYG@tok@m\endcsname{\def\PYG@tc##1{\textcolor[rgb]{0.40,0.40,0.40}{##1}}}
\expandafter\def\csname PYG@tok@gh\endcsname{\let\PYG@bf=\textbf\def\PYG@tc##1{\textcolor[rgb]{0.00,0.00,0.50}{##1}}}
\expandafter\def\csname PYG@tok@gu\endcsname{\let\PYG@bf=\textbf\def\PYG@tc##1{\textcolor[rgb]{0.50,0.00,0.50}{##1}}}
\expandafter\def\csname PYG@tok@gd\endcsname{\def\PYG@tc##1{\textcolor[rgb]{0.63,0.00,0.00}{##1}}}
\expandafter\def\csname PYG@tok@gi\endcsname{\def\PYG@tc##1{\textcolor[rgb]{0.00,0.63,0.00}{##1}}}
\expandafter\def\csname PYG@tok@gr\endcsname{\def\PYG@tc##1{\textcolor[rgb]{1.00,0.00,0.00}{##1}}}
\expandafter\def\csname PYG@tok@ge\endcsname{\let\PYG@it=\textit}
\expandafter\def\csname PYG@tok@gs\endcsname{\let\PYG@bf=\textbf}
\expandafter\def\csname PYG@tok@gp\endcsname{\let\PYG@bf=\textbf\def\PYG@tc##1{\textcolor[rgb]{0.00,0.00,0.50}{##1}}}
\expandafter\def\csname PYG@tok@go\endcsname{\def\PYG@tc##1{\textcolor[rgb]{0.53,0.53,0.53}{##1}}}
\expandafter\def\csname PYG@tok@gt\endcsname{\def\PYG@tc##1{\textcolor[rgb]{0.00,0.27,0.87}{##1}}}
\expandafter\def\csname PYG@tok@err\endcsname{\def\PYG@bc##1{\setlength{\fboxsep}{0pt}\fcolorbox[rgb]{1.00,0.00,0.00}{1,1,1}{\strut ##1}}}
\expandafter\def\csname PYG@tok@kc\endcsname{\let\PYG@bf=\textbf\def\PYG@tc##1{\textcolor[rgb]{0.00,0.50,0.00}{##1}}}
\expandafter\def\csname PYG@tok@kd\endcsname{\let\PYG@bf=\textbf\def\PYG@tc##1{\textcolor[rgb]{0.00,0.50,0.00}{##1}}}
\expandafter\def\csname PYG@tok@kn\endcsname{\let\PYG@bf=\textbf\def\PYG@tc##1{\textcolor[rgb]{0.00,0.50,0.00}{##1}}}
\expandafter\def\csname PYG@tok@kr\endcsname{\let\PYG@bf=\textbf\def\PYG@tc##1{\textcolor[rgb]{0.00,0.50,0.00}{##1}}}
\expandafter\def\csname PYG@tok@bp\endcsname{\def\PYG@tc##1{\textcolor[rgb]{0.00,0.50,0.00}{##1}}}
\expandafter\def\csname PYG@tok@fm\endcsname{\def\PYG@tc##1{\textcolor[rgb]{0.00,0.00,1.00}{##1}}}
\expandafter\def\csname PYG@tok@vc\endcsname{\def\PYG@tc##1{\textcolor[rgb]{0.10,0.09,0.49}{##1}}}
\expandafter\def\csname PYG@tok@vg\endcsname{\def\PYG@tc##1{\textcolor[rgb]{0.10,0.09,0.49}{##1}}}
\expandafter\def\csname PYG@tok@vi\endcsname{\def\PYG@tc##1{\textcolor[rgb]{0.10,0.09,0.49}{##1}}}
\expandafter\def\csname PYG@tok@vm\endcsname{\def\PYG@tc##1{\textcolor[rgb]{0.10,0.09,0.49}{##1}}}
\expandafter\def\csname PYG@tok@sa\endcsname{\def\PYG@tc##1{\textcolor[rgb]{0.73,0.13,0.13}{##1}}}
\expandafter\def\csname PYG@tok@sb\endcsname{\def\PYG@tc##1{\textcolor[rgb]{0.73,0.13,0.13}{##1}}}
\expandafter\def\csname PYG@tok@sc\endcsname{\def\PYG@tc##1{\textcolor[rgb]{0.73,0.13,0.13}{##1}}}
\expandafter\def\csname PYG@tok@dl\endcsname{\def\PYG@tc##1{\textcolor[rgb]{0.73,0.13,0.13}{##1}}}
\expandafter\def\csname PYG@tok@s2\endcsname{\def\PYG@tc##1{\textcolor[rgb]{0.73,0.13,0.13}{##1}}}
\expandafter\def\csname PYG@tok@sh\endcsname{\def\PYG@tc##1{\textcolor[rgb]{0.73,0.13,0.13}{##1}}}
\expandafter\def\csname PYG@tok@s1\endcsname{\def\PYG@tc##1{\textcolor[rgb]{0.73,0.13,0.13}{##1}}}
\expandafter\def\csname PYG@tok@mb\endcsname{\def\PYG@tc##1{\textcolor[rgb]{0.40,0.40,0.40}{##1}}}
\expandafter\def\csname PYG@tok@mf\endcsname{\def\PYG@tc##1{\textcolor[rgb]{0.40,0.40,0.40}{##1}}}
\expandafter\def\csname PYG@tok@mh\endcsname{\def\PYG@tc##1{\textcolor[rgb]{0.40,0.40,0.40}{##1}}}
\expandafter\def\csname PYG@tok@mi\endcsname{\def\PYG@tc##1{\textcolor[rgb]{0.40,0.40,0.40}{##1}}}
\expandafter\def\csname PYG@tok@il\endcsname{\def\PYG@tc##1{\textcolor[rgb]{0.40,0.40,0.40}{##1}}}
\expandafter\def\csname PYG@tok@mo\endcsname{\def\PYG@tc##1{\textcolor[rgb]{0.40,0.40,0.40}{##1}}}
\expandafter\def\csname PYG@tok@ch\endcsname{\let\PYG@it=\textit\def\PYG@tc##1{\textcolor[rgb]{0.25,0.50,0.50}{##1}}}
\expandafter\def\csname PYG@tok@cm\endcsname{\let\PYG@it=\textit\def\PYG@tc##1{\textcolor[rgb]{0.25,0.50,0.50}{##1}}}
\expandafter\def\csname PYG@tok@cpf\endcsname{\let\PYG@it=\textit\def\PYG@tc##1{\textcolor[rgb]{0.25,0.50,0.50}{##1}}}
\expandafter\def\csname PYG@tok@c1\endcsname{\let\PYG@it=\textit\def\PYG@tc##1{\textcolor[rgb]{0.25,0.50,0.50}{##1}}}
\expandafter\def\csname PYG@tok@cs\endcsname{\let\PYG@it=\textit\def\PYG@tc##1{\textcolor[rgb]{0.25,0.50,0.50}{##1}}}

\def\PYGZus{\char`\_}

\def\PYGZsh{\char`\#}

\def\PYGZsq{\char`\'}

\makeatother

\makeatletter
\def\PYGborland@reset{\let\PYGborland@it=\relax \let\PYGborland@bf=\relax%
    \let\PYGborland@ul=\relax \let\PYGborland@tc=\relax%
    \let\PYGborland@bc=\relax \let\PYGborland@ff=\relax}
\def\PYGborland@tok#1{\csname PYGborland@tok@#1\endcsname}
\def\PYGborland@toks#1+{\ifx\relax#1\empty\else%
    \PYGborland@tok{#1}\expandafter\PYGborland@toks\fi}
\def\PYGborland@do#1{\PYGborland@bc{\PYGborland@tc{\PYGborland@ul{%
    \PYGborland@it{\PYGborland@bf{\PYGborland@ff{#1}}}}}}}
\def\PYGborland#1#2{\PYGborland@reset\PYGborland@toks#1+\relax+\PYGborland@do{#2}}

\expandafter\def\csname PYGborland@tok@w\endcsname{\def\PYGborland@tc##1{\textcolor[rgb]{0.73,0.73,0.73}{##1}}}
\expandafter\def\csname PYGborland@tok@c\endcsname{\let\PYGborland@it=\textit\def\PYGborland@tc##1{\textcolor[rgb]{0.00,0.53,0.00}{##1}}}
\expandafter\def\csname PYGborland@tok@cp\endcsname{\def\PYGborland@tc##1{\textcolor[rgb]{0.00,0.50,0.50}{##1}}}
\expandafter\def\csname PYGborland@tok@cs\endcsname{\let\PYGborland@bf=\textbf\def\PYGborland@tc##1{\textcolor[rgb]{0.00,0.53,0.00}{##1}}}
\expandafter\def\csname PYGborland@tok@s\endcsname{\def\PYGborland@tc##1{\textcolor[rgb]{0.00,0.00,1.00}{##1}}}
\expandafter\def\csname PYGborland@tok@sc\endcsname{\def\PYGborland@tc##1{\textcolor[rgb]{0.50,0.00,0.50}{##1}}}
\expandafter\def\csname PYGborland@tok@m\endcsname{\def\PYGborland@tc##1{\textcolor[rgb]{0.00,0.00,1.00}{##1}}}
\expandafter\def\csname PYGborland@tok@k\endcsname{\let\PYGborland@bf=\textbf\def\PYGborland@tc##1{\textcolor[rgb]{0.00,0.00,0.50}{##1}}}
\expandafter\def\csname PYGborland@tok@ow\endcsname{\let\PYGborland@bf=\textbf}
\expandafter\def\csname PYGborland@tok@nt\endcsname{\let\PYGborland@bf=\textbf\def\PYGborland@tc##1{\textcolor[rgb]{0.00,0.00,0.50}{##1}}}
\expandafter\def\csname PYGborland@tok@na\endcsname{\def\PYGborland@tc##1{\textcolor[rgb]{1.00,0.00,0.00}{##1}}}
\expandafter\def\csname PYGborland@tok@gh\endcsname{\def\PYGborland@tc##1{\textcolor[rgb]{0.60,0.60,0.60}{##1}}}
\expandafter\def\csname PYGborland@tok@gu\endcsname{\def\PYGborland@tc##1{\textcolor[rgb]{0.67,0.67,0.67}{##1}}}
\expandafter\def\csname PYGborland@tok@gd\endcsname{\def\PYGborland@tc##1{\textcolor[rgb]{0.00,0.00,0.00}{##1}}\def\PYGborland@bc##1{\setlength{\fboxsep}{0pt}\colorbox[rgb]{1.00,0.87,0.87}{\strut ##1}}}
\expandafter\def\csname PYGborland@tok@gi\endcsname{\def\PYGborland@tc##1{\textcolor[rgb]{0.00,0.00,0.00}{##1}}\def\PYGborland@bc##1{\setlength{\fboxsep}{0pt}\colorbox[rgb]{0.87,1.00,0.87}{\strut ##1}}}
\expandafter\def\csname PYGborland@tok@gr\endcsname{\def\PYGborland@tc##1{\textcolor[rgb]{0.67,0.00,0.00}{##1}}}
\expandafter\def\csname PYGborland@tok@ge\endcsname{\let\PYGborland@it=\textit}
\expandafter\def\csname PYGborland@tok@gs\endcsname{\let\PYGborland@bf=\textbf}
\expandafter\def\csname PYGborland@tok@gp\endcsname{\def\PYGborland@tc##1{\textcolor[rgb]{0.33,0.33,0.33}{##1}}}
\expandafter\def\csname PYGborland@tok@go\endcsname{\def\PYGborland@tc##1{\textcolor[rgb]{0.53,0.53,0.53}{##1}}}
\expandafter\def\csname PYGborland@tok@gt\endcsname{\def\PYGborland@tc##1{\textcolor[rgb]{0.67,0.00,0.00}{##1}}}
\expandafter\def\csname PYGborland@tok@err\endcsname{\def\PYGborland@tc##1{\textcolor[rgb]{0.65,0.09,0.09}{##1}}\def\PYGborland@bc##1{\setlength{\fboxsep}{0pt}\colorbox[rgb]{0.89,0.82,0.82}{\strut ##1}}}
\expandafter\def\csname PYGborland@tok@kc\endcsname{\let\PYGborland@bf=\textbf\def\PYGborland@tc##1{\textcolor[rgb]{0.00,0.00,0.50}{##1}}}
\expandafter\def\csname PYGborland@tok@kd\endcsname{\let\PYGborland@bf=\textbf\def\PYGborland@tc##1{\textcolor[rgb]{0.00,0.00,0.50}{##1}}}
\expandafter\def\csname PYGborland@tok@kn\endcsname{\let\PYGborland@bf=\textbf\def\PYGborland@tc##1{\textcolor[rgb]{0.00,0.00,0.50}{##1}}}
\expandafter\def\csname PYGborland@tok@kp\endcsname{\let\PYGborland@bf=\textbf\def\PYGborland@tc##1{\textcolor[rgb]{0.00,0.00,0.50}{##1}}}
\expandafter\def\csname PYGborland@tok@kr\endcsname{\let\PYGborland@bf=\textbf\def\PYGborland@tc##1{\textcolor[rgb]{0.00,0.00,0.50}{##1}}}
\expandafter\def\csname PYGborland@tok@kt\endcsname{\let\PYGborland@bf=\textbf\def\PYGborland@tc##1{\textcolor[rgb]{0.00,0.00,0.50}{##1}}}
\expandafter\def\csname PYGborland@tok@sa\endcsname{\def\PYGborland@tc##1{\textcolor[rgb]{0.00,0.00,1.00}{##1}}}
\expandafter\def\csname PYGborland@tok@sb\endcsname{\def\PYGborland@tc##1{\textcolor[rgb]{0.00,0.00,1.00}{##1}}}
\expandafter\def\csname PYGborland@tok@dl\endcsname{\def\PYGborland@tc##1{\textcolor[rgb]{0.00,0.00,1.00}{##1}}}
\expandafter\def\csname PYGborland@tok@sd\endcsname{\def\PYGborland@tc##1{\textcolor[rgb]{0.00,0.00,1.00}{##1}}}
\expandafter\def\csname PYGborland@tok@s2\endcsname{\def\PYGborland@tc##1{\textcolor[rgb]{0.00,0.00,1.00}{##1}}}
\expandafter\def\csname PYGborland@tok@se\endcsname{\def\PYGborland@tc##1{\textcolor[rgb]{0.00,0.00,1.00}{##1}}}
\expandafter\def\csname PYGborland@tok@sh\endcsname{\def\PYGborland@tc##1{\textcolor[rgb]{0.00,0.00,1.00}{##1}}}
\expandafter\def\csname PYGborland@tok@si\endcsname{\def\PYGborland@tc##1{\textcolor[rgb]{0.00,0.00,1.00}{##1}}}
\expandafter\def\csname PYGborland@tok@sx\endcsname{\def\PYGborland@tc##1{\textcolor[rgb]{0.00,0.00,1.00}{##1}}}
\expandafter\def\csname PYGborland@tok@sr\endcsname{\def\PYGborland@tc##1{\textcolor[rgb]{0.00,0.00,1.00}{##1}}}
\expandafter\def\csname PYGborland@tok@s1\endcsname{\def\PYGborland@tc##1{\textcolor[rgb]{0.00,0.00,1.00}{##1}}}
\expandafter\def\csname PYGborland@tok@ss\endcsname{\def\PYGborland@tc##1{\textcolor[rgb]{0.00,0.00,1.00}{##1}}}
\expandafter\def\csname PYGborland@tok@mb\endcsname{\def\PYGborland@tc##1{\textcolor[rgb]{0.00,0.00,1.00}{##1}}}
\expandafter\def\csname PYGborland@tok@mf\endcsname{\def\PYGborland@tc##1{\textcolor[rgb]{0.00,0.00,1.00}{##1}}}
\expandafter\def\csname PYGborland@tok@mh\endcsname{\def\PYGborland@tc##1{\textcolor[rgb]{0.00,0.00,1.00}{##1}}}
\expandafter\def\csname PYGborland@tok@mi\endcsname{\def\PYGborland@tc##1{\textcolor[rgb]{0.00,0.00,1.00}{##1}}}
\expandafter\def\csname PYGborland@tok@il\endcsname{\def\PYGborland@tc##1{\textcolor[rgb]{0.00,0.00,1.00}{##1}}}
\expandafter\def\csname PYGborland@tok@mo\endcsname{\def\PYGborland@tc##1{\textcolor[rgb]{0.00,0.00,1.00}{##1}}}
\expandafter\def\csname PYGborland@tok@ch\endcsname{\let\PYGborland@it=\textit\def\PYGborland@tc##1{\textcolor[rgb]{0.00,0.53,0.00}{##1}}}
\expandafter\def\csname PYGborland@tok@cm\endcsname{\let\PYGborland@it=\textit\def\PYGborland@tc##1{\textcolor[rgb]{0.00,0.53,0.00}{##1}}}
\expandafter\def\csname PYGborland@tok@cpf\endcsname{\let\PYGborland@it=\textit\def\PYGborland@tc##1{\textcolor[rgb]{0.00,0.53,0.00}{##1}}}
\expandafter\def\csname PYGborland@tok@c1\endcsname{\let\PYGborland@it=\textit\def\PYGborland@tc##1{\textcolor[rgb]{0.00,0.53,0.00}{##1}}}

\makeatother

\usepackage[english]{babel}  
\addto\extrasenglish{

}

\newcommand{\Ex}[0]{\ensuremath{\mathbb{E}}}

\DeclareMathOperator*{\argmax}{arg\,max}

\newcommand{\nblink}[1]{\href{https://github.com/McWilliamsCenter/deep_galaxy_models/blob/master/#1.ipynb}{\faFileCodeO}}
\newcommand{\github}{\href{https://github.com/McWilliamsCenter/deep_galaxy_models}{\faGithub}}

\title[Generative galaxy model]{Deep Generative Models for Galaxy Image Simulations}

\author[Lanusse et al.]{Fran\c{c}ois Lanusse,$^{1}$\thanks{E-mail: francois.lanusse@cea.fr}
Rachel Mandelbaum,$^{2}$
Siamak Ravanbakhsh,$^{3,4}$
\newauthor
Chun-Liang Li,$^{5}$
Peter Freeman,$^{6}$
and Barnab\'as P\'oczos$^{5}$
\\
$^{1}$AIM, CEA, CNRS, Universit\'e Paris-Saclay, Universit\'e Paris Diderot, Sorbonne Paris Cit\'e, F-91191 Gif-sur-Yvette, France\\
$^{2}$McWilliams Center for Cosmology, Department of Physics, Carnegie Mellon University, Pittsburgh, PA 15213, USA\\
$^{3}$School of Computer Science, McGill University, Montreal, QC, Canada\\
$^{4}$Mila, Quebec AI Institute, Montreal, QC, Canada\\
$^{5}$School of Computer Science, Carnegie Mellon University, Pittsburgh, PA 15213, USA\\
$^{6}$Department of Statistics \& Data Science, Carnegie Mellon University, Pittsburgh, PA 15213, USA\\
}
\date{Accepted XXX. Received YYY; in original form ZZZ}

\pubyear{2020}

\begin{document}
\label{firstpage}
\pagerange{\pageref{firstpage}--\pageref{lastpage}}
\maketitle

\begin{abstract}
Image simulations are essential tools for preparing and validating the analysis of current and future wide-field optical surveys. However, the galaxy models used as the basis for these simulations are typically limited to simple parametric light profiles, or use a fairly limited amount of available space-based data. In this work, we propose a methodology based on Deep Generative Models to create complex models of galaxy morphologies that may meet the image simulation needs of upcoming surveys. We address the technical challenges associated with learning this morphology model from noisy and PSF-convolved images by building a hybrid Deep Learning/physical Bayesian hierarchical model for observed images, explicitly accounting for the Point Spread Function and  noise properties. The generative model is further made conditional on physical galaxy parameters, to allow for  sampling  new light profiles from specific galaxy populations. We demonstrate our ability to train and sample from such a model on galaxy postage stamps from the  HST/ACS  COSMOS survey, and validate the quality of the model using a range of second- and higher-order  morphology statistics. Using this set of statistics, we demonstrate significantly more realistic morphologies using these deep generative models compared to conventional parametric models. To help make these generative models practical tools for  the community, we introduce \texttt{GalSim-Hub}, a community-driven repository of generative models, and a framework for incorporating generative models within the \texttt{GalSim} image simulation software. \github
\end{abstract}

\begin{keywords}
methods: statistical -- techniques: image processing
\end{keywords}



\section{Introduction}

Image simulations are fundamental tools for the analysis of modern wide-field optical surveys.  For example, they play a crucial role in estimating and calibrating systematic biases in weak lensing analyses \citep[e.g.,][]{Fenech2017, Samuroff2017,Mandelbaum2018}. In preparation for upcoming missions, major collaborations including the Rubin Observatory Legacy Survey of Space and Time (LSST) Dark Energy Science Collaboration \footnote{\url{https://lsstdesc.org/}} \citep[DESC;][]{DESC2012}, the Euclid Consortium\footnote{\url{https://www.euclid-ec.org/}} \citep{Laureijs2011}, and the Roman Space Telescope \footnote{\url{https://roman.gsfc.nasa.gov/}} \citep{Spergel2015}, are currently in the process of generating large scale image simulations of their respective surveys \citep[e.g.][]{Sanchez2020, Troxel2019}.

Despite the importance of these large simulation campaigns, the most common approach to simulating galaxy light profiles is to rely on simple analytic profiles such as S\'ersic profiles \citep[e.g.][]{Kacprzak2019, Kannawadi2019}. Besides their simplicity, the main motivation for this choice is the existence of prescriptions for the distribution of the parameters of these profiles.  These distributions can be directly drawn from observations by fitting these profiles to existing surveys such as COSMOS \citep{Mandelbaum2012,Griffith2012}, or provided by empirical \citep{Korytov2019} or Semi-Analytic Models (SAM)s \citep{Somerville2015}. These simple models therefore may be used as the basis for fairly realistic image simulations, with galaxies at least matching the correct size and ellipticity distributions as a function of magnitude and redshift.

However, as the precision of modern surveys increases, so does the risk of introducing model biases from these simple assumptions on galaxy light profiles. The impact of model bias for weak lensing shape measurement was for instance explicitly investigated in \cite{Mandelbaum2015}, and the impact of galaxy morphologies was measurable, if subdominant, in the calibration of the HSC Y1 shape catalog \citep{Mandelbaum2018}.  Beyond their direct effect on shape measurement, assumptions  about galaxy light profiles impact various stages of  the upstream data reduction pipeline, and in particular the deblending step. It is for instance expected that a majority of galaxies  observed  by LSST will be blended  with their neighbors, given that blending impacts $\sim 60$\% of galaxies in the similar wide survey of  the Hyper Suprime Cam \citep[HSC;][]{Bosch2017}. As current deblenders, like \texttt{Scarlet} \citep{Melchior2018}, rely on simple assumptions of monotonicity and symmetry of  galaxy light profiles, having access to simulations with non-trivial galaxy light profiles will be essential to properly assess systematic deblender-induced biases in number counts, galaxy photometry, and other properties.

Several works have explored galaxy models going beyond simple parametric light profiles. One of the simplest extension is the inclusion of a so-called random knots component \citep{Zhang2015a, Sheldon2017}, constituted of point sources randomly distributed along the galaxy light profile, which can model knots of star formation. However, building a realistic prescription for the parameters of this knots component (number of point sources, flux, spatial distribution) is not trivial. In newer large-scale image simulations produced by the LSST DESC (DESC Collaboration, \textit{in prep.}) a model for this component was obtained by fitting a three component (bulge+disk+knots) light profiles to HST COSMOS image, and then used in image simulations. Similarly, a prescription for how to place these knots based on fitting nearby galaxies was proposed in \citet{Plazas2019}. \citet{Massey2004} built a generative model for deep galaxy images based on a shapelet representation, generating new galaxies by perturbing the shapelet decomposition of galaxies fitted in a training set. Finally, image simulations can be based on existing deep imaging, either directly \citep[e.g.][]{Mandelbaum2012, Mandelbaum2018}, or after denoising \citep{Maturi2017} to simulate deeper observations.

With the recent  advent of Deep Learning, several works have investigated the use  of deep generative  models to learn galaxy morphologies.  In pioneering work, \cite{Regier2015a} proposed the use of Variational AutoEncoders \citep[VAE;][]{Kingma2013} as tools to model galaxy images. The use of VAEs and the first use of Generative Adversarial Networks \citep[GAN;][]{Goodfellow2014} for astronomical images was further explored in \citet{Ravanbakhsh2017a}, along with conditional image generation. More recently, \cite{Fussell2019} demonstrated  an application of a StackGan model \citep{Zhang2016} to generate high-resolution images from the Galaxy Zoo 2 SDSS sample \citep{Willett2013}. Similarly, the generation of large galaxy fields using GANs was demonstrated in \citet{Smith2019}. Beyond generic image simulations, GANs and VAEs have also been proposed to address complex tasks dependent on galaxy morphologies when processing astronomical images, such as deblending \citep{Reiman2019, Arcelin2020} or deconvolution \citep{Schawinski2017a}. Very recently, \cite{Lanusse2019} proposed to use likelihood-based generative models (e.g. PixelCNN++; \citealt{Salimans2017}) as priors for solving astronomical inverse problems such as deblending within a physically motivated Bayesian framework. 

\begin{figure*}
    \centering
    \includegraphics[width=\textwidth]{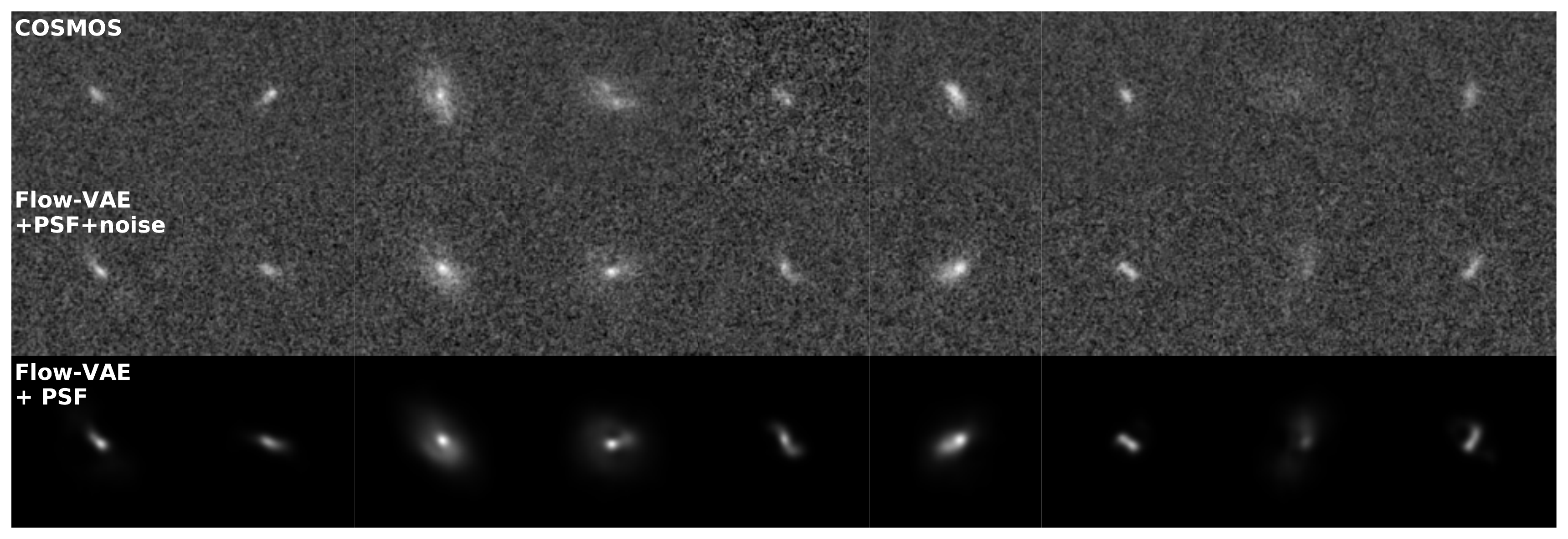}
    \caption{Samples from real COSMOS galaxies (top), and \textbf{random} draws from the generative model (middle) with matching PSF and noise, and \textbf{conditioned} on the size, magnitude, and redshift of the corresponding COSMOS galaxy. The bottom row shows the same generated light-profiles but without observational noise. Because 
    of this conditioning, generated galaxies (middle) are consistent in appearance with the corresponding COSMOS galaxy. \nblink{Figure1}}
    \label{fig:demo}
\end{figure*}

All these precursor works have demonstrated the great potential of Deep Learning techniques, but none of them have gone beyond the stage of simple proof of principle. The goal of this paper is to provide the tools needed to build generative models from astronomical data  in practice, i.e., accounting for the instrumental response and  observing conditions, as well as providing the software framework to make these tools easily usable by the community as part of the broadly used \texttt{GalSim}\footnote{\url{https://github.com/GalSim-developers/GalSim}} image simulation software \citep{Rowe2015}.

To this end, we demonstrate how latent variable models such as GANs and VAEs can be embedded as part of a broader Bayesian hierarchical model, providing a physical model for the Point Spread Function (PSF) and noise properties of individual observations. This view of the problem allows us in principle to learn a denoised and PSF-deconvolved model for galaxy morphology, from data acquired under various observing conditions, and even different instruments.  A  variety of  deep generative models can be used under this framework. As a specific example we propose here a model based on a VAE,  complemented by a latent-space normalizing flow \citep{Dinh2016,Rezende2015} to achieve high sample quality. We call this hybrid model a Flow-VAE. We further make our proposed generative model conditional on physical galaxy properties (e.g., magnitude, size, etc) which allows us to sample specific galaxy populations. This is a crucial element to be  able to connect image generation to mock galaxy catalogs for generating survey images from a simulated  extragalactic object  catalog.
We train our proposed generative model on a sample of galaxies from the HST/ACS COSMOS survey, and evaluate the realism of the generated images  under different morphology statistics that include, but go beyond, the second moments, including  size, ellipticity, Gini, M20, and MID statistics \citep{Freeman2013}. Overall, we  find excellent agreement between the generated images and real COSMOS images under these statistics and demonstrate that these mock galaxies are quantitatively more complex than simple parametric profiles.

Finally, we introduce \texttt{GalSim-Hub}\footnote{\url{https://github.com/McWilliamsCenter/galsim_hub}},  a library and repository of trained generative models, interfaced directly into \texttt{GalSim}, with the hope that  the  availability of  such tools will foster the development of generative models of even higher quality, as well as  a broader access to these  methods by  the  community. All the tools used to train the generative models presented in this work rely on the \texttt{Galaxy2Galaxy}\footnote{\url{https://github.com/ml4astro/galaxy2galaxy}} framework (Lanusse et al, {\em in prep.}).

\bigskip

After stating the problem of learning from heterogeneous data  in \autoref{sec:learningcorrupted}, we  introduce our proposed generative  model, dubbed Flow-VAE, in \autoref{sec:flowvae}. We  train this model and thoroughly evaluate its performance  in \autoref{sec:cosmosVAE}. A summary of our results and future prospects for this work are discussed  in \autoref{sec:conclusion}.

\section{Learning from corrupted data}
\label{sec:learningcorrupted}

While most of the Deep Learning literature on generative models is concerned with natural images (photographic pictures of daily life scenes), learning generative models for galaxy light profiles from astronomical images requires specific technical challenges to be addressed.  These include properly dealing with the noise in the observations as well as accounting for the PSF. The question we will focus on in this section is how to learn a noise-free and PSF-deconvolved distribution of galaxy morphologies, from data acquired under varying observing conditions, or  even from different instruments. This can be done  by \textit{complexifying the causal structure of GANs and VAEs}\footnote{Credit to this expression and underlying idea goes to David W. Hogg.}, or in other words, integrating these deep generative models as part of a larger Hierarchical Bayesian Model allowing us to cleanly combine these Deep Learning elements within a physically motivated model of the data. In the end, our goal is to produce results like those shown on \autoref{fig:demo} where the deep generative model only learns galaxy morphologies, while PSF and noise can be added explicitly for a specific instrument or survey.  A very similar idea, but for forward modeling multiband photometry instead of images, was proposed in \cite{Leistedt2019}. A machine learning component modeling Spectral Energy Distribution (SED) templates was embedded in a larger physical and causal hierarchical model of galaxy photometry, in order to jointly constrain SED templates and photometric redshifts.

\subsection{Latent Variable Models as components in larger physically motivated Bayesian networks}

\begin{figure}
    \centering
    \includegraphics{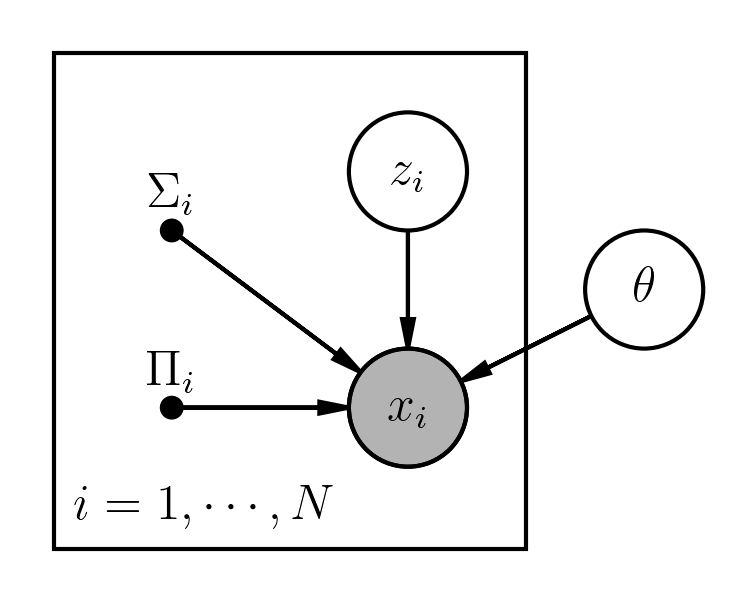}
    \caption{Probabilistic graphical model for observed galaxy images. For each galaxy $i$, the pixel values  $x_i$ are obtained by transforming an input random variable $z_i$ through a parametric generator function $g_\theta(z_i)$ before applying the instrumental PSF $\Pi_i$ and adding Gaussian noise with covariance $\Sigma_i$. \nblink{Figure_pgm}}
    \label{fig:pgm}
\end{figure}

In this work we consider deep latent variable models (LVM), describing a target distribution $p(x)$ in terms of a latent variable $\bm{z}$ drawn from a prior distribution $p(z)$ and mapped into data space by a parametric function $g_\theta$, usually referred to as the \textit{generator} and taking the form in practice of a deep neural network. While they differ on other points, both VAEs and GANs fall under this class of models. These LVMs can be thought of as flexible parametric models to represent otherwise unknown distributions. As such they can be readily integrated in wider Bayesian networks to fill in parts of the graphical model for which we do not have an explicit formulation.

Let us consider the specific problem of modeling observed galaxy images, with pixel values $\bm{x}$. Making explicit use  of our  knowledge of the PSF and noise properties of the image, we can model  these pixel intensities as being related to the actual galaxy light profile $I$ through:
\begin{equation}
	\bm{x }_i = \Pi_i \ast I_i + \bm{n}_i \;,
\end{equation}
where  $\Pi$ represents  the  PSF (accounting for telescope optics, atmospheric perturbation, and the pixel response of the sensor) and $n$ describes observational noise. In this model, the PSF can typically be estimated from the images of stars in the data itself by the pipeline, or retrieved from a physical optical model of the instrument. Similarly, while the  specific noise realization $n$ is unknown, its statistical properties can also be estimated separately from photon-counting expectations or empirical statistics in the imaging. In this work, we will assume a Gaussian noise model, with pixel covariance matrix $\Sigma_i$. Note that this covariance can be non-diagonal as the result of the warping of images during data processing. With those two components under control, only the galaxy light profile $I$ remains without a tractable physical model; this is where we can introduce a LVM.

Let  us assume that any galaxy light profile $I$ can be realized by a LVM mapping a latent variable $\bm{z}$ into an image through a generator function  $I_i=g_\theta(\bm{z}_i)$.  We can now describe the pixel values of an image as:
\begin{equation}
	\bm{x}_i = \Pi_i \ast g_\theta(\bm{z}_i) + \bm{n}_i \;.
\end{equation}
Note that while $\bm{x}_i, \bm{z}_i, \Pi_i. \bm{n}_i$ are specific to each observation, the parameters $\theta$ of the LVM are not. A graphical representation of  this model is provided in \autoref{fig:pgm}.

Learning a model for galaxy morphology now amounts to finding a set of parameters $\theta_\star$ for the generator $g_\theta$ that ensures that the empirical distribution of the data $p(\bm{x})$ is consistent with the distribution $p_\theta( \bm{x})$ described by this Bayesian Hierarchical Model:
\begin{equation}
	p_\theta( \bm{x}) = \prod\limits_{i=1}^N  \int p_\theta(\bm{x}_i | \Pi_i, \Sigma_i, \bm{z}_i) p(\bm{z}_i) \mathrm{d}\bm{z}_i  \;.
\end{equation}
Solving the optimization problem involved in finding the parameters $\theta_\star$ is typically a difficult task due to the marginalization over latent variables $\bm{z}$ involved in this expression. Both VAEs and GANs provide tractable solutions, although they differ in methodology: GANs are likelihood-free methods, i.e., they bypass the need to evaluate the marginalized likelihood $p_{\theta}(\bm{x})$ and instead only require the ability to sample from  it. On the other hand, VAEs rely on the existence a tractable variational lower bound to the  marginalized likelihood. 

\subsection{Modeling the data likelihood}

In this work, we assume the observational noise to be Gaussian distributed, with pixel covariance $\mathbf{\Sigma}$ and 0 mean. This is a  common  model for sky subtracted images where the noise coming from the dark current and the Poisson fluctuations of the sky background and galaxy can reliably be modeled as Gaussian distributed. 

In many situations of interests, $\mathbf{\Sigma}$ is assumed to be diagonal in pixel space and potentially spatially varying. In this case, the likelihood  of the data can  conveniently be expressed in pixel space as:
\begin{multline}
    	\log p_{\theta} (\bm{x}_i | \Pi_i, \Sigma_i, \bm{z}_i) = - \frac{1}{2} (\bm{x}_i - \Pi_i \ast g_{\theta}(\bm{z}_i))^t \mathbf{\Sigma}_i^{-1} (\bm{x}_i -  \Pi_i \ast g_{\theta}(\bm{z}_i)) \\ \ + \ cst \;.
	\label{eq:likellihood_pixel}
\end{multline}

Alternatively, if the noise is known to be correlated but stationary, another tractable assumption is to assume the noise covariance to be diagonal in Fourier space. 
\begin{multline}
	\log p_{\theta} (\bm{x}_i | \Pi_i, \Sigma_i, \bm{z}_i) = \\ - \frac{1}{2} \mathcal{F}\left(\bm{x}_i -  \Pi_i \ast  g_{\theta}(\bm{z}_i)\right)^t \mathbf{\Sigma}_i^{-1} \mathcal{F}\left(\bm{x}_i -  \Pi_i \ast g_{\theta}(\bm{z}_i) \right)  \ + \ cst \;,
\end{multline}
where $\mathcal{F}$ is the forward Fourier transform.

In implicit models such as GANs, evaluating the likelihood is not required; all that is needed is the  ability to sample from it. This can be achieved by adding Gaussian noise to the PSF-convolved images created by the generator, before sending them to the discriminator.
Note that this step is particularly crucial for GAN generation of noisy images, as there is not enough entropy in the input latent space of the GAN to generate an independent noise realization of the size of an image, needed to match the noise in the data. We find that in practice, without adding noise samples, the generator tries to learn some noise patterns that are actually replicated from image to image.

\section{Learning the Generative Model by Variational Inference}
\label{sec:flowvae}

In this section, we briefly introduce the various Deep Learning notions underlying the generative models proposed in this work.

\subsection{Auto-Encoding Variational Bayes}

\begin{figure}
\resizebox{\columnwidth}{!}{%
\begin{tikzpicture}[auto, ultra thick, node distance=1.2cm, 
			   layer/.style={draw, rectangle, ultra thick, minimum height=6cm},
			   code/.style={draw, rectangle,  ultra thick, minimum height=3cm, fill=lightgray},
			   condition/.style={draw, rectangle,  ultra thick, minimum height=1.3cm, fill=lightgray}]

			\node[inner sep=0pt] (input) {\includegraphics[width=3cm]{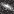}};
			\node[above=0.25cm  of input ] (input_legend) {\Large $x$};
			\node[condition, below = 0.2cm of input]	(cond0) {};
			\node[below= 0.25cm  of cond0] (cond_text) {\Large $y$};

			\node [layer, right = of input] (p1) {}; 
			\node [layer, right = 0.5cm of p1] (p2) {}; 
			\node [layer, right = 0.5cm of p2] (p3) {}; 
			\node[above= 0.5cm  of p2](p2_text) {\Large $q_{\phi}(z \mid x, y)$};
			\draw [  decoration={ brace, mirror,  raise=0.5cm  },  decorate]  (p1.south west) -- (p3.south east)  node [pos=0.5,anchor=north,yshift=-0.55cm, text width=4cm, align=center] {\Large Inference network}; 

			\node[code, right = of p3]	(code) {};
			\node[above= 0.5cm  of code] (code_text) {\Large $z \sim q_\phi(z | x, y)$};;

			\node [layer, right = of code] (q1) {}; 
			\node [layer, right = 0.5cm of q1] (q2) {}; 
			\node [layer, right = 0.5cm of q2] (q3) {}; 
			\node[above= 0.5cm  of q2](q2_text) {\Large $p_{\theta}(x \mid z)$};
			\draw [decoration={ brace, mirror,  raise=0.5cm},  decorate]  (q1.south west) -- (q3.south east)  node [pos=0.5,anchor=north,yshift=-0.55cm, text width=4cm, align=center] {\Large Generator network}; 
			
			\node[inner sep=0pt, right = of q3] (output) {\includegraphics[width=3cm]{fig/forward_G3-crop_img21}};
			\node[above= 0.5cm of output](output_text) {\Large $x^\prime \sim p_{\theta}(x \mid z)$};
			
			\draw [->] (input) -- (p1); 
			\draw[->] (cond0) -- (cond0-|p1.west);
			\draw [->] (p1) -- (p2); 
			\draw [->] (p2) -- (p3); 
			
			\draw [->] (p3) -- (code);
			
			\draw [->] (code) -- (q1);
			\draw [->] (q1) -- (q2);
			\draw [->] (q2) -- (q3);
			
			\draw [->] (q3) -- (output);
\end{tikzpicture}}
\caption{Schematic representation of a Variational Auto-Encoder. The inference network $q_\phi(z|x,y)$ is tasked with predicting the  posterior distribution of a given image $x$ and additional information $y$ in latent space $z$. Access to this posterior distribution allows for efficient training of the generative model $p_\theta(x | z)$, which models the pixel-level image given the latent variable $z$.}
\label{fig:vae}
\end{figure}

Auto-Encoding Variational Bayes (AEVB), also known as the Variational Auto-Encoder, is a framework introduced in \cite{Kingma2013} to enable tractable maximum likelihood inference of the parameters of a directed graphical model with continuous latent variables. In such models, one assumes that the observations $\bm{x}$ are generated following a random process involving unobserved latent variables $\bm{z}$ according to some parametric distribution $p_{\bm{\theta}}(\bm{x}, \bm{z}) = p_{\bm{\theta}}(\bm{x} | \bm{z}) p(\bm{z})$, where $\bm{\theta}$ are parameters of this distribution which we aim to adjust so that the marginal distribution $p_{\bm{\theta}}(\bm{x})$ matches closely the empirical distribution of the data. In the context of the model presented in the previous section, these parameters $\bm{\theta}$ will correspond to the weights and biases of the neural network $g_{\theta}$ introduced to model galaxy light profiles.

In this model, we have the freedom to choose any parametric distribution $p_{\bm{\theta}}(\bm{x} | \bm{z})$ to describe the mapping between latent and data space; we only ask for it to be sufficiently flexible to effectively represent the data, and to be easy to sample from. A natural choice is to assume a given parametric likelihood function for the data, and use deep neural networks to learn the mapping from latent space to these distribution parameters. As an example, assuming a Gaussian likelihood for the data, the expression of $p_{\bm{\theta}}(\bm{x} | \bm{z})$ becomes:
\begin{equation}
	p_{\bm{\theta}}(\bm{x} | \bm{z}) = \mathcal{N}(\mu_\theta(\bm{z}), \Sigma_\theta(\bm{z})) \;,
\end{equation}
where $\mu_\theta$ and $\Sigma_\theta$ can be deep neural networks depending on a set of parameters $\bm{\theta}$. Training such a model now involves adjusting these parameters as to maximize the marginal likelihood of the model:
\begin{equation}
 \hat{\bm{\theta}} = \argmax_{\bm{\theta}} p_{\theta} (\bm{x}) =  \argmax_{\bm{\theta}} \int p_{\theta}(\bm{x} | \bm{z}) p(\bm{z}) d \bm{z} \;.
\end{equation}
What makes this problem difficult however is that evaluating this marginal likelihood, or its derivatives with respect to the parameters $\bm{\theta}$, is typically intractable analytically and too costly using Monte Carlo techniques. 

The idea behind AEVB is to introduce an inference model $q_{\bm{\varphi}}(\bm{z} | \bm{x})$ to estimate for each example $\bm{x}$ the true posterior density $p_{\bm{\theta}}(\bm{z}| \bm{x})$ in the latent space. This model, also known as the recognition model, is performing approximate posterior inference, typically by using a deep neural network to predict the parameters of a parametric distribution (e.g. $q_{\bm{\varphi}} =  \mathcal{N}(\mu_\varphi(\bm{x}), \sigma^2_\varphi(\bm{x}))$). This model is essentially replacing a costly MCMC by a single call to a deep neural network to approximate $p_{\bm{\theta}}(\bm{z}|\bm{x})$, this is known as \textit{amortized} variational inference. The usefulness of this inference model becomes clear when deriving the Kullback-Leibler divergence between this approximation and the true posteriors:
\begin{align}
	\mathbb{D}_{KL} [ q_{\phi}(\bm{z} | \bm{x}) || p(\bm{z} | \bm{x})] &= \Ex_{q} [ \log q(\bm{z} | \bm{x}) - \log p(\bm{z} | \bm{x}) ] \nonumber\\
													   &= \Ex_{q} [\log q(\bm{z} | \bm{x})  - \log p(\bm{z})]+ \log p(\bm{x})  \nonumber \\
													   &\quad - \Ex_q[\log p(\bm{x} | \bm{z})]  \nonumber \\
													   & = \mathbb{D}_{KL}[q_{\phi}(\bm{z} | \bm{x}) || p(\bm{z})] +\log p(\bm{x}) \nonumber \\
													   & \quad - \Ex_q[\log p(\bm{x} | \bm{z})] \nonumber \;.
\end{align}
Reordering the terms of this expression leads to:
\begin{multline}
 \log p(\bm{x}) = \Ex_q[\log p_{\bm{\theta}}(\bm{x} | \bm{z})] - \mathbb{D}_{KL}[q_{\phi}(\bm{z} | \bm{x}) || p(\bm{z})] \\ + \underbrace{\mathbb{D}_{KL} [ q_{\phi}(\bm{z} | \bm{x}) || p(\bm{z} | \bm{x})]}_{ \geq 0} \;.
 \label{eq:divergence}
\end{multline}
Taking into account the fact that the KL divergence is always positive, this leads to the following lower bound on the marginal log likelihood of $x$, known as the Evidence Lower Bound (ELBO):
\begin{equation}
	 \log p(\bm{x}) \geq \Ex_{z \sim q_{\varphi}(. | x)}[\log p_{\bm{\theta}}(\bm{x} | \bm{z})]  -  \mathbb{D}_{KL}[q_{\varphi}(\bm{z} | \bm{x}) || p(\bm{z})] \;.
	 \label{eq:ELBO}
\end{equation} 
Contrary to the original marginal likelihood, the ELBO is now completely tractable, as neither $p(\bm{x})$ or $p(\bm{z}|\bm{x})$ appear in the rhs. The final key element of AEVB is a stochastic gradient descent algorithm (using the so-called reparametrization trick) to efficiently optimize this lower bound in practice \citep{Kingma2013}.

This combination of a recognition and generative model, illustrated by \autoref{fig:vae}, is reminiscent of traditional auto-encoders, which follow the same idea of compressing the information down to a latent space and reconstructing the input signal from this low dimensional representation. The difference comes from the second term in the ELBO in Eq.~\eqref{eq:ELBO} which prevents the latent space representation of particular examples from collapsing to a delta function, and instead promotes the representation learned by the model to stay close to the specified prior $p(\bm{z})$.

Despite the satisfying mathematical motivation for the VAE, it is known that this model usually leads to overly smooth images.  The reasons for this problem are an active field of research in machine learning, but are likely due to the difficulty of performing accurate amortized inference of the posterior while training the generator \citep{Kingma2016a, Cremer2018a, He2019}. In this work, instead of trying to address the sub-optimalities of the variational inference, we follow a different direction, originally proposed in \citet{Engel2017}, which is to relax the KL divergence term in Eq.~\eqref{eq:ELBO}, and introducing a second model for modeling the latent space aggregate posterior.  

\subsection{VAE with free bits}

Empirically, it is known that training a VAE will tend to find a solution that conforms to the prior $p(\bm{z})$ at the expense of reconstruction and sample quality, leading to over-regularized solutions. A number of different approaches have been proposed to force the model towards better optimization minima \citep{Sonderby2016}, in particular the idea of starting the optimization without the KL divergence term in the ELBO and slowly increasing its strength during training. Rather than relying on an annealing scheme,  \cite{Kingma2016a} proposed to allow for some amount of information to be communicated through the bottleneck of the auto-encoder without being penalized by the KL divergence:
\begin{equation}
	 \mathcal{L}_{\lambda} = \Ex_{z \sim q_{\varphi}(. | x)}[\log p_{\bm{\theta}}(\bm{x} | \bm{z})]  - \max\left( \lambda, \ \mathbb{D}_{KL}[q_{\varphi}(\bm{z} | \bm{x}) || p(\bm{z})] \right) \;.
	 \label{eq:ELBO_lambda}
\end{equation} 
The $\lambda$ parameter controls how many \textit{free bits} of information can be used by the model before incurring an actual penalty. Allowing for more free bits leads to better reconstruction quality as more information about the input image is being transferred to the generator, but allowing for too many free bits essentially removes the regularization of the latent space and we recover a conventional autoencoder, from which we cannot directly sample as the aggregated posterior no longer has any incentive to follow the prior. 

The approach proposed in \cite{Engel2017} is to significantly down-weight the KL divergence term in the ELBO, so as to emphasize reconstruction quality first and foremost, at the cost of less regularization in the latent space. Images sampled from this model with Gaussian prior appear significantly distorted and usually meaningless. As  a  solution to that problem, the authors propose to learn a separate model that models a so-called \textit{realism constraint}, essentially learning to sample from the aggregate posterior of the data as opposed to the prior. This approach leads to both sharp images and high quality samples, on par with different methods such as GANs can generate. An additional benefit of this approach is that the VAE can be trained once, while the actual posterior sampling model can always be refined later and even made conditional, without needing to retrain the entire auto-encoder (which is in general more costly).

We follow a similar approach in this work, reducing the latent space regularization of the VAE by using the ELBO with free bits loss function defined  in Eq.~\eqref{eq:ELBO_lambda}. In the next section we will introduce a second latent space model to learn how to sample realistic images.

\subsection{Flow-VAE: Learning the VAE posterior distribution}
\label{sec:NF}

The quality of VAE samples depends strongly on how successful the model is at matching the aggregate posterior distribution of the data to the prior. If this posterior departs from the prior, sampling from the prior will not lead to good quality samples matching the data distribution of the training set. Such failures in matching the posterior to prior may naturally arise in VAEs when the latent space regularization is weaker than the data fidelity term. Another common situation is when training a Conditional VAE, where the model is incentivized to decorrelate the latent variables from the conditional variables \citep[e.g.][]{Ravanbakhsh2017a}. This is never perfect, and again the data posterior never completely matches the Gaussian prior and usually exhibits some residual correlations with the conditional variables.

To alleviate these issues, a solution is to train an additional latent space model to learn the aggregate posterior of the data for a given trained VAE. This model can also be made conditional so that it can allow to sample conditionally the latent variables. This two-step process has the advantage of decoupling the training of the VAE on actual images, which can be costly, from the training of the latent-space sampling model, which is much typically much faster. This means for instance that once a VAE is trained, it is possible to inexpensively build a number of conditional models, simply by training different conditional sampling models.

While \cite{Engel2017} proposed to use a GAN to model the latent space, we adopt instead a normalizing flow, a type of Neural Density Estimation method with exact log likelihood, which achieves state-of-the-art results in density estimation while being significantly more stable than GANs. Furthermore, normalizing flows are not susceptible to \textit{mode collapse} \citep[e.g.][]{Salimans2016,Che2016}, a common failure mode of GANs which translates into a lack of variety in generated samples. Normalizing flows model a target distribution in terms of a parametric bijective mapping $g_\theta$ from a prior distribution $p(\bm{z})$ to the target distribution $p(\bm{x})$. Under this model, the probability of a sample $\bm{x}$ from the data set can be computed by applying a change of variable formula: 
\begin{equation}
	p_\theta(\bm{x}) = p(\bm{z}) \left| \frac{\partial g_\theta}{\partial \bm{z}} \right| (\bm{z}) \quad \mbox{ with } \bm{z} = g_\theta^{-1}(\bm{x}) \;.
\end{equation}
With this explicit expression for the likelihood of a data sample under the model, fitting the normalizing flow can be done by minimizing the negative log likelihood of the data:
\begin{equation}
\mathcal{L} = -  \log p_\theta(\bm{x}) = -  \log p(\bm{z})  - \log\left| \frac{\partial g_\theta}{\partial \bm{z}} \right| (\bm{z}) \;.
\end{equation}
Under the assumption that $\mathbb{D}_{KL}(p || p_\theta ) =0$ is actually attainable (i.e., that $p_\theta$, and hence $g_\theta$, is flexible enough), it will be achieved at the minimum of this loss function.  

The main practical challenge in building  normalizing flows is in designing a mapping $g_\theta$ that needs to be both bijective, and with a tractable Jacobian determinant. One such possible efficient design is the  Masked Autoregressive Flow (MAF) introduced in \cite{Papamakarios2017}. A MAF layer is defined by the following mapping:
\begin{equation}
    g_\theta(\bm{x}) = \sigma_\theta(\bm{x}) \odot \bm{x} + \mu_\theta(\bm{x}) \;,
\end{equation}
where $\odot$ is the Hadamard product (element-wise multiplication), and $\sigma_\theta$ and $\mu_\theta$ are autoregressive functions, i.e. the $i$th dimension of the output $[\mu_\theta(x_1, \ldots, x_N)]_i$ only depends on the previous dimensions $(x_1, \ldots, x_{i-1})$. These autoregressive functions are implemented in practice using a masked dense neural network, as proposed in \cite{Germain2015}. Given the autoregressive nature of this mapping, its Jacobian takes on a simple lower triangular structure, which makes computing its determinant simple: 
\begin{equation}
   \log \left| \frac{\partial g_\theta(\bm{x})}{\partial \bm{x}}  \right| = \sum_{i=0}^{N} \log\sigma_{\theta, i}(\bm{x}) \; .
\end{equation} 
While a single layer of a MAF cannot model very complex mappings, more expressive models can be obtained by chaining several flow layers:
\begin{equation}
    g_\theta(\bm{x}) = f_\theta^0 \circ f_\theta^{1} \circ \ldots \circ f_\theta^{N} (\bm{x}) \;.
\end{equation}

In this work, we further extend the baseline MAF model to build a conditional density estimator $p_\theta (\bm{x} | \bm{y})$. This can be achieved by providing the conditional variable as an input of the shift and scaling functions $\sigma_\theta$ and $\mu_\theta$ so that $z_i = f(\bm{y}, x_0, \ldots, x_{i-1})$. The resulting conditional density estimator can be used to learn the latent aggregate posterior of the VAE, conditioned on particular parameters of interest, for instance galaxy size or brightness.

\begin{figure}
    \centering
    \includegraphics[width=\columnwidth]{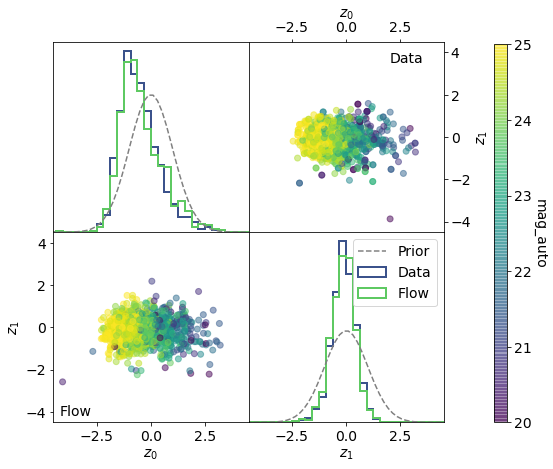}
    \caption{Illustration of  latent variable $\mathbf{z}$ distribution as a function of galaxy size for auto-encoded  galaxies (top right) and samples from the latent normalizing flow (bottom left).  As can be seen from the upper corner plot, the 2d histograms of the latent variables for auto-encoded galaxies can significantly depart from the assumed isotropic Gaussian prior (dashed grey lines) used during training of the VAE. We can also see strong correlations between latent variables $\textbf{z}$ and properties such the magnitude. Both of these effects, i.e. departures from Gaussianity and magnitude-dependence are successfully modeled by the latent normalizing flow in the bottom corner plots.  \nblink{Figure_ConditionalFlow} }
    \label{fig:posterior_sampling}
\end{figure}
The upper right corner of \autoref{fig:posterior_sampling} illustrates the first two dimensions of the empirical aggregate posterior distribution of a VAE with a 16-d latent space (detailed in \autoref{sec:vae_model}). The color indicates the i-band magnitude of the galaxy corresponding to each encoded point. As can be seen from this example, not only is the posterior distribution significantly non-Gaussian, it also exhibits a clear and non-trivial dependence on the galaxy magnitude. The bottom left part of \autoref{fig:posterior_sampling} illustrates samples from a conditional normalizing flow which not only reproduces correctly the overall posterior distribution, but also captures the correct dependency on magnitude.

\section{Generative Model trained on COSMOS} 
\label{sec:cosmosVAE}

In this  section we present our reference model for the GalSim COSMOS sample using the Flow-VAE approach introduced above.

\subsection{The GalSim COSMOS Sample}

Our training set is based on the COSMOS HST Advanced Camera for Surveys (ACS) field \citep{Koekemoer2007,Scoville2007b,Scoville2007}, a 1.64 deg$^2$ contiguous survey acquired with the ACS Wide Field Channel, through the F814W filter (``Broad I''). Based on this survey, a dataset of deblended galaxy postage stamps \citep{Leauthaud2007, Mandelbaum2012} was compiled as part of the GREAT3 challenge \citep{Mandelbaum2014}, and forms the basis for our training set. The processing steps and selection criteria required to build this sample are introduced in \cite{Mandelbaum2012} and we direct the interested reader to the \textit{Real Galaxy Dataset} appendix of \cite{Mandelbaum2014} for a comprehensive description of this sample. We use the deep F814W$<25.2$ version of the dataset, provided with the \texttt{GalSim} simulation software \citep{Rowe2015} through the \texttt{COSMOSCatalog} class, which provides in addition for each postage stamps the HST PSF, the noise power spectrum, and a  set of galaxy properties (e.g., size, magnitude, photometric redshift). As discussed further in the next section, among these additional parameters, we will in particular make use of the Source Extractor F814W magnitude \texttt{mag\_auto}, the (PSF-deconvolved) half-light radius \texttt{hlr}, and photometric redshift \texttt{zphot} fields. Applying the default quality cut of \texttt{exclusion\_level='marginal'} with the COSMOSCatalog leaves us with a sample of 81,500 galaxy postage stamps, which we divide into training and testing sets containing 80,000 and 1,500 galaxies, respectively.

For training, we draw these galaxies at the original  0.03$^{\prime\prime}$/pix resolution of the coadded images, on postage stamps of size $128\times128$, convolved with their original PSF and using noise padding. For each galaxy, we also store an image of the associated PSF and noise power spectrum. An illustration of these postage stamps is provided on the top row of \autoref{fig:demo}.

\subsection{Generative model}
\label{sec:vae_model}

\subsubsection{VAE Architecture and  Training}

For the VAE, we adopt a deep ResNet architecture, based on seven stages of downsampling, with each stage composed on two residual blocs. The depth after a first channel-wise dense embedding layer is set to 16, and is multiplied by two at each downsampling step until reaching a maximum depth of 512. After these purely convolutional layers, we compress the latent representation down to a vector of 16 dimensions using a single dense layer, outputting the mean and standard deviation for a mean-field Gaussian posterior model $q_\phi(z | x)$. Likewise,  the 16-d latent representation is decoded back to the input dimension of the convolutional generator using  a single dense layer. The rest of the generative model is mirroring  the architecture of the encoder.
At the final layer of the generator, we apply a softplus\footnote{softplus activation: $f(x)=\ln(1+\exp(x))$} activation function to ensure the positivity of the light profile generated by the model.

As explained in \autoref{sec:learningcorrupted}, the output of the VAE is then convolved with the known PSF of the input image, and the likelihood that enters the ELBO in \autoref{eq:ELBO} is
computed using the known noise covariance. The results presented here are obtained using a diagonal approximation to the covariance (i.e. using \autoref{eq:likellihood_pixel}) as it is simpler than a non-diagonal covariance and yields very comparable results. 
In order to very slightly regularize the pre-convolved light profile generated by the VAE and prevent non-physical very high frequency we include in the loss function, in addition to the ELBO, a small Total Variation\footnote{TV: $\ell_1$ norm of the gradients of the image, $TV(x) = \parallel \nabla x \parallel_1 $} (TV) term that penalizes these potential high frequency artifacts which are not otherwise constrained by the data. We add this TV term to the loss with a factor 0.01, which makes it very subdominant to the rest of the loss function.

In addition, in order to provide the encoder network with all the information it needs to learn a deconvolved galaxy image, we feed it an image of the PSF for each example, in addition to galaxy image presented at the input. Without this additional information, the model wouldn't be able to perform the desired inference task.

\begin{table}
	\centering
	\caption{Hyper-parameters used to train the VAE model.}
	\label{tab:simulation}
	\begin{tabular}{ll}
		\hline
		\hline
		Parameter & Value \\
		\hline
		\textit{Architecture choices} & \\
		Number of ResNet blocks & 7  (for each encoder/decoder)\\
		Input depth & 16 \\
		Maximum depth & 512 \\
		Bottleneck size & 16 \\		
		& \\
		\textit{Optimizer and Training} & \\
		Optimizer & Adafactor \\
		Number of iterations & 125,000\\
		Base learning rate & 0.001 \\
		Learning schedule & square root decay\\
		 Batch size & 64\\
		 Free-bits & 4\\
		 Total Variation factor & 0.01\\
		\hline
	\end{tabular}
\end{table}

Training of the model is performed using Adafactor \citep{Shazeer2018}, a variant of the popular ADAM optimizer \citep{Kingma2015} with an adaptive learning rate, with parameters described in \autoref{tab:simulation}. Note that to make training of this deep encoder/decoder model more efficient, we use the following two strategies:
\begin{itemize}
	\item[-] Similarly to a UNet \citep{Ronneberger2015}, we allow for transverse connections between corresponding stages of the encoder/decoder during training. Concretely, a random sub-sample of the feature maps at a given level of the generator are simply duplicated from the encoder to the decoder, thus short-circuiting part of the model. This allows the last layers of the generator to start training, even though the deeper layers are not correctly trained yet. This fraction of random duplication of the encoder feature maps to the decoder is slowly decreased  during training, until these transverse connections are fully removed.	
	\item[-] To help the dense bottleneck layers to learn a mapping close to the identity,  we add an $\ell_2$ penalty between inputs and outputs of the bottleneck. The  strength of  this penalty is again slowly decreased during training. 
\end{itemize}

\subsubsection{Latent Normalizing Flow training}

Once the auto-encoder is trained, we reuse the encoder with fixed weights to generate samples from the aggregate posterior of the training set images. These samples of the latent space variable $\mathbf{z}$ are in turn used to train the latent space Normalizing Flow  described in \autoref{sec:NF}. This model relies on 8 layers of MAF stages, each of these stages is using two masked dense layers of size 256. Between MAF stages, we alternate between performing a random shuffling of all dimensions and reversing the order of the tensor dimensions, so as to facilitate the mixing between dimensions. Each of the MAF stages is using both shift and scale operations. To help improve the stability of the model during training, we further apply clipping to the output scaling coefficients $\sigma_\theta(x)$ generated by each MAF layers. To improve conditional modeling, the additional features $\bm{y}$ are standardized by removing their means and scaling their standard deviation to one. 

Training is performed with the ADAM optimizer over 50,000 iterations with a base learning rate of 0.001, following a root square decay with number of iterations.

\bigskip

The trained model is available on \texttt{GalSim-Hub} under the model name ``hub:Lanusse2020''. We direct the interested reader to \autoref{sec:galsimhub} for an example of how to use this model with \texttt{GalSim}.

\subsection{Auto-Encoding Verification}

\begin{figure}
	\includegraphics[width=\columnwidth]{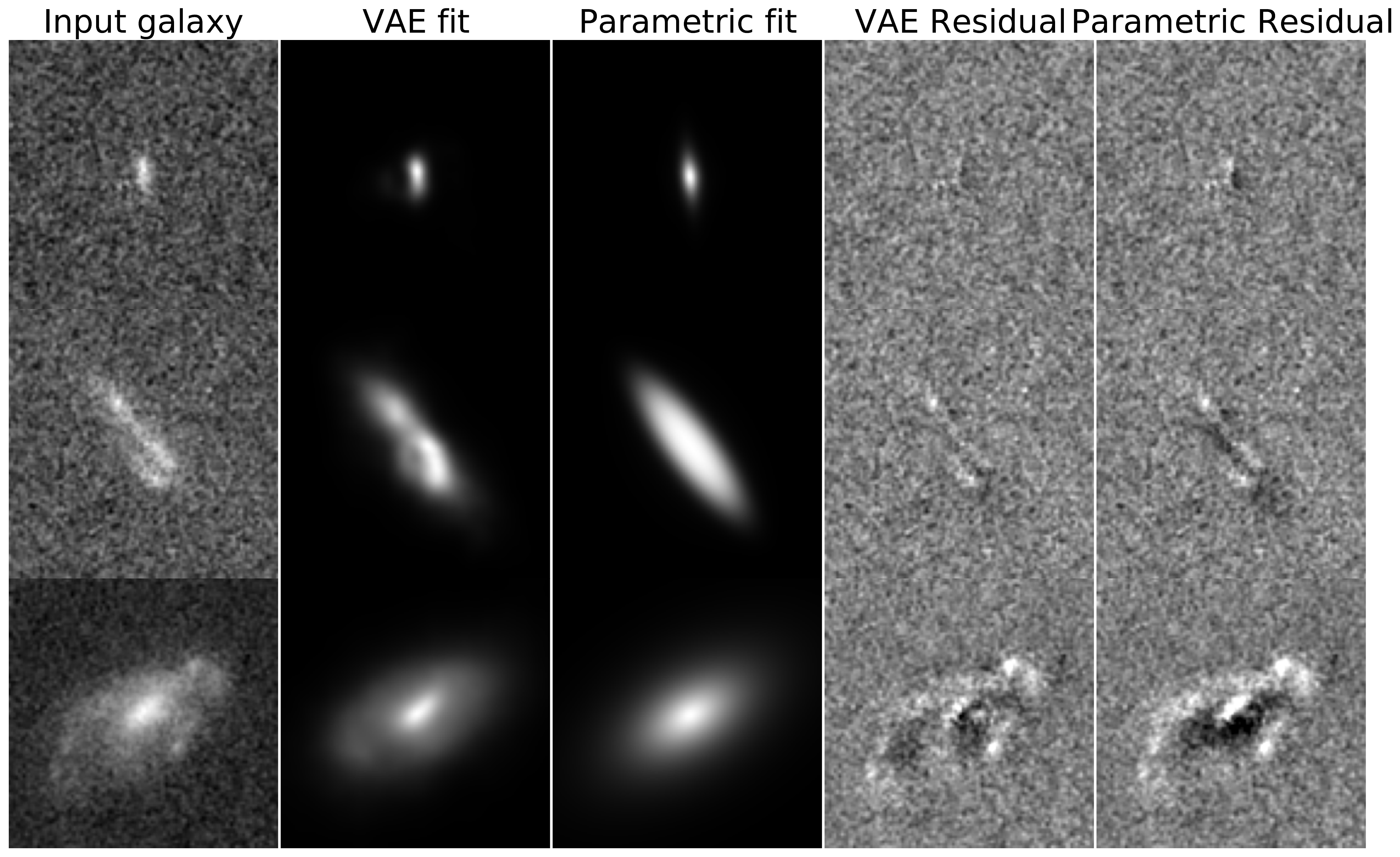}
	\caption{Reconstruction of input images (first column) by the VAE (second column) and by Parametric fit (third column). Residuals for both VAE and parametric models are shown on fourth and fifth columns respectively. From top to bottom are illustrated representative objects of increasing size; smaller compact objects (top) are accurately reconstructed by the model, while larger galaxies exhibit some modeling residuals (bottom). Note that the VAE models are always more complex than their parametric counterparts.  \nblink{Figure_autoencode}}
	\label{fig:autoencode}
\end{figure}
Before testing the quality of the full generative model, we first assess the representation power of the VAE model on galaxies from the testing set. \autoref{fig:autoencode} is illustrating how galaxies of different sizes are auto-encoded by the VAE model, compared to a conventional parametric fit to these light profiles (described in the next section).  As can be seen, smaller galaxies are very well modeled by the auto-encoder, but for larger galaxies, the model exhibits smoother light profiles, illustrating one of the limitations of such an autoencoder model. The free bits of information used during training of the VAE are intended to mitigate that effect,  but are only partially successful. We note furthermore that these large galaxies are under-represented in the training sample, meaning that the model is comparatively less incentivized to correctly model these bright and large galaxies compared to smaller and fainter objects. Accounting and compensating for this training set imbalance could be an avenue to alleviate this effect, but at the price of changing the galaxy distribution being modeled by the VAE.

In all cases, we see on the two rightmost columns of \autoref{fig:autoencode} that VAE residuals are significantly smaller than the  residuals of the best parametric fit, indicating a better modeling. It is also worth highlighting that in the VAE case, these fitted light profiles are parameterized by only 16 numbers obtained in a single pass of amortized inference, and yet yield more accurate results than the iterative parametric fitting. 

\subsection{Sample generation validation}

In this section, we quantitatively assess the quality of the light profiles generated by our models in terms of several summary statistics, including second order 
moments and morphological image statistics specifically designed to identify non-smooth and non-monotonic light profiles \citep{Freeman2013}.

To perform these comparisons, we generate three different samples:
\begin{itemize}
	\item \texttt{COSMOS sample}: Real HST COSMOS galaxies, drawn from the GalSim real galaxy sample.
	\item \texttt{Parametric sample}: Parametric galaxies drawn from the GalSim best parametric model of real COSMOS galaxies, either single S\'ersic or a Bulge+Disk model depending  on the best fitting model.
	\item \texttt{Mock sample}: Artificial galaxies drawn from the generative model, conditioned on the magnitude, size, and redshift of real COSMOS galaxies.
\end{itemize} 
Each tuple of galaxies from these three sets is drawn with the same PSF and matching noise properties as to allow direct comparison. 

\subsubsection{Second-order moments}

We first evaluate the quality of the model in terms of second-order moments of the light profile, defined as:
\begin{equation}
	Q_{i,j} = \frac{ \int \mathrm{d}^2 x I(x) W(x) x_i x_j}{ \int \mathrm{d}^2 x I(x) W(x) } \;,
\end{equation}
where $I$ is the light profile, $W$ is a weighting function, and $x_i,x_j$ are centroid-subtracted pixel coordinates. This centroid is in practice adaptively estimated from the image itself. We rely on the \texttt{GalSim} HSM module, which implements adaptive moment estimation \citep{Bernstein2002, Hirata2003a} of the PSF-convolved, elliptical Gaussian-weighted second moments.

Based on these measured moments $\mathbf{Q}$, we use the determinant radius $\sigma = \det \mathbf{Q} ^{1/4}$ to characterize the size of galaxies, and we also consider their ellipticity $g$ defined as:
\begin{equation}
    g = g_1 + i g_2 = \frac{Q_{1,1} - Q_{2,2} - 2i Q_{1,2}}{Q_{1,1} + Q_{2,2} + 2 (Q_{1,1} Q_{2,2} - Q_{1,2}^2)^{1/2}} \;.
\end{equation}
Note that this definition is distinct from the alternative \textit{distortion} definition of a galaxy ellipticity.

\begin{figure}
    \begin{subfigure}[b]{1.0\columnwidth}
		\includegraphics[width=\columnwidth]{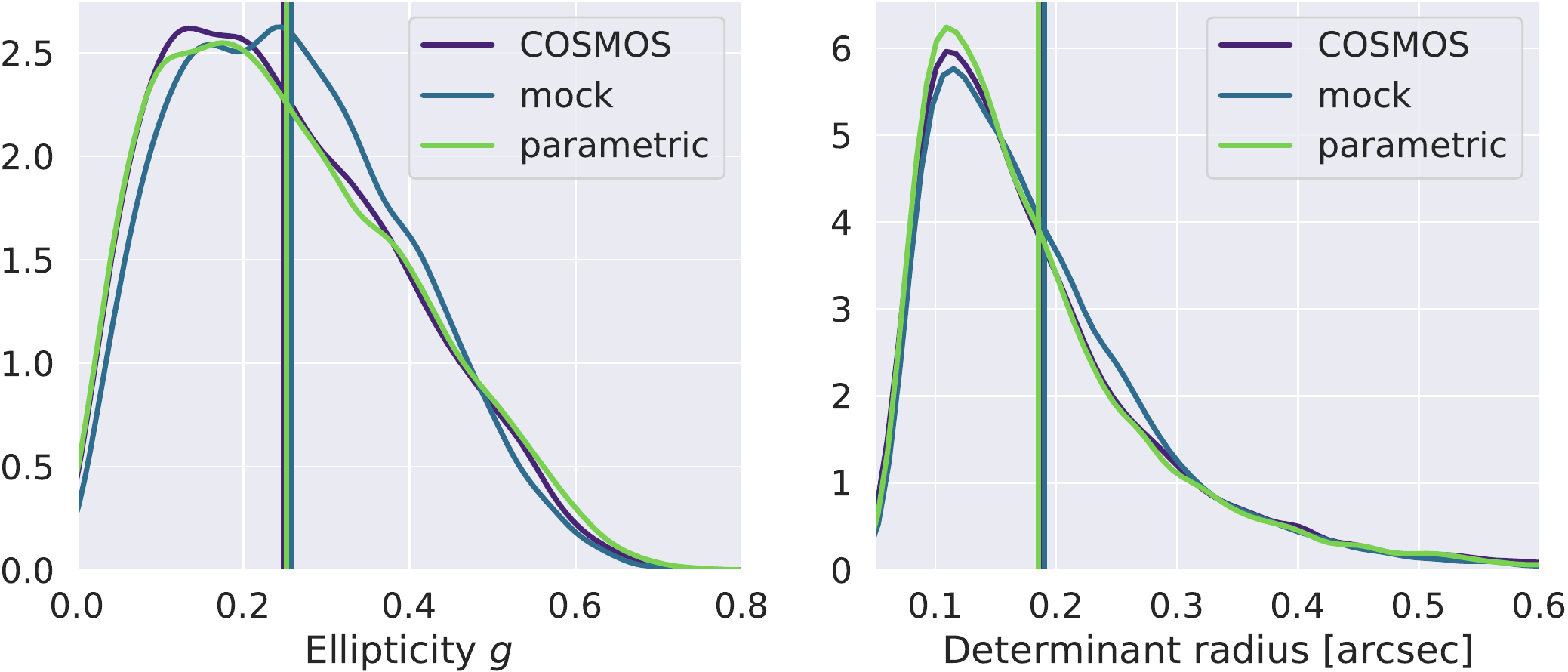}
		\caption{Marginal distributions}
	\end{subfigure}\\
	\begin{subfigure}[b]{1.0\columnwidth}
		\includegraphics[width=0.97\columnwidth]{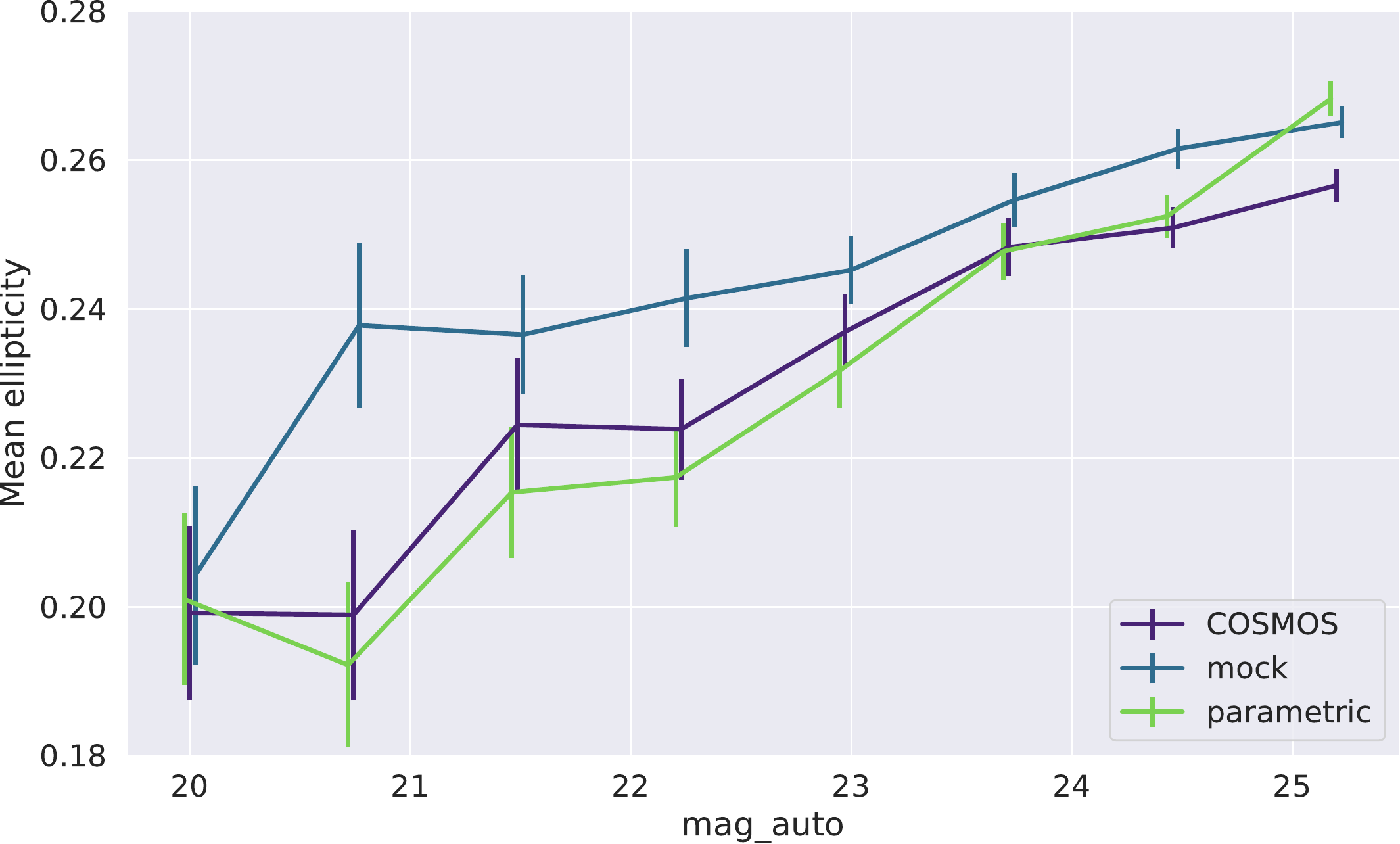}
		\caption{Ellipticity as a function of  magnitude}
	\end{subfigure}
	\caption{Comparison of second-order moments between COSMOS galaxies, parametric fits, and VAE samples. The vertical lines in (a) indicate the means of the respective distributions. The error bars in (b) indicate the 1-$\sigma$ error on the mean ellipticity. \nblink{Figure_Moments}}
	\label{fig:moments}
\end{figure}

\autoref{fig:moments} compares the marginal distribution of determinant radius $\sigma$ and ellipticity $|g|$ for the three different samples. We find excellent agreement between the reference COSMOS distribution and galaxies generated from the generative model, with a 4\% difference in mean ellipticity and 1\% difference in mean size.

\begin{figure*}
	\includegraphics[width=0.9\columnwidth]{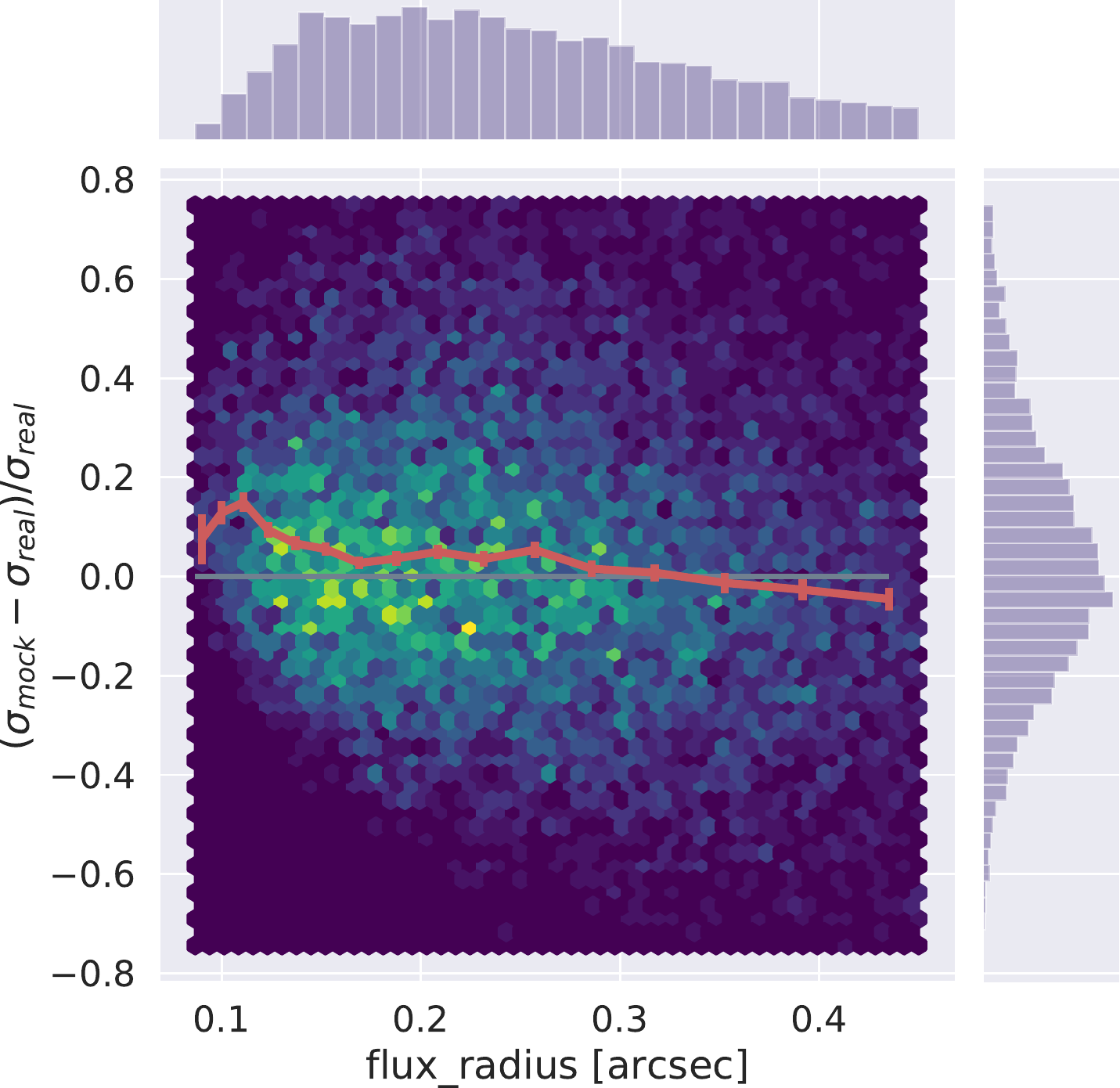} \quad
	\includegraphics[width=0.9\columnwidth]{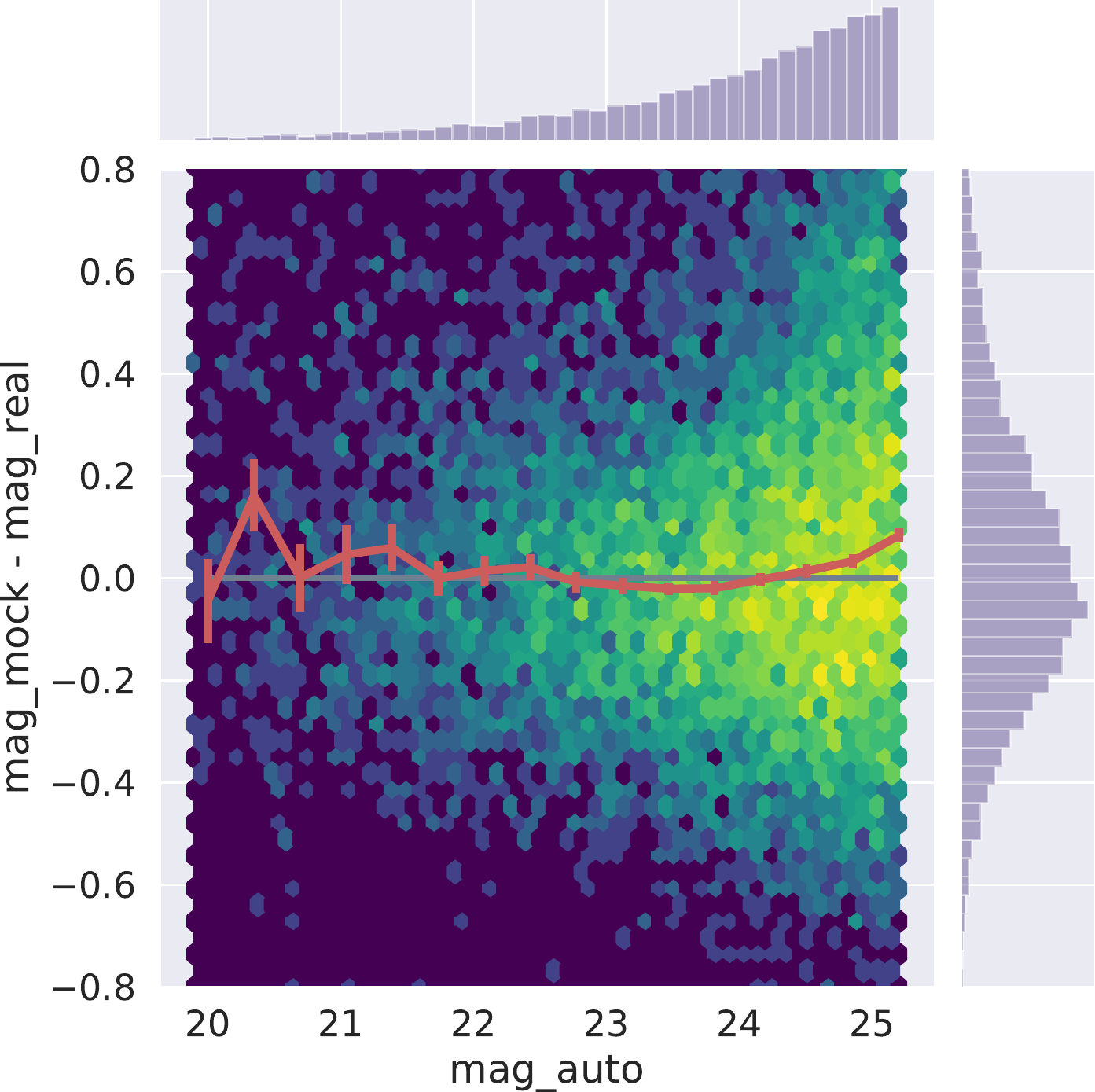}
	\caption{Comparison of measured determinant radius  and magnitude between pairs of COSMOS galaxies and VAE samples, as a function of half light radius and magnitude of the real galaxy, which are also used to condition the corresponding VAE samples. The solid red line represents the median of the difference in size and flux, in bins of the corresponding conditional quantities. The error bars indicate the $1-\sigma$ uncertainty on that median value. \nblink{Figure_Conditioning}
	}
	\label{fig:conditional_sampling}
\end{figure*}

In addition to comparing the overall distribution of size and ellipticity, we can test the quality of the  conditional sampling with \autoref{fig:conditional_sampling} showing for each pair of real and mock galaxy the difference in size and flux, as a function of the corresponding conditional variable. The red line in these plots shows the median of the corresponding residual distribution in bins of size and  magnitude. On these simple statistics, we find that the conditioning is largely unbiased, but note an overall $\sim 27\%$ scatter in size, and $\sim 0.3$ in magnitude. For these two properties however,  while the conditioning is not extremely precise, a desired size and flux can always be imposed after sampling from the generative model, using GalSim light profile manipulation utilities.

\subsubsection{Morphological statistics}

\begin{figure}
	\includegraphics[width=\columnwidth]{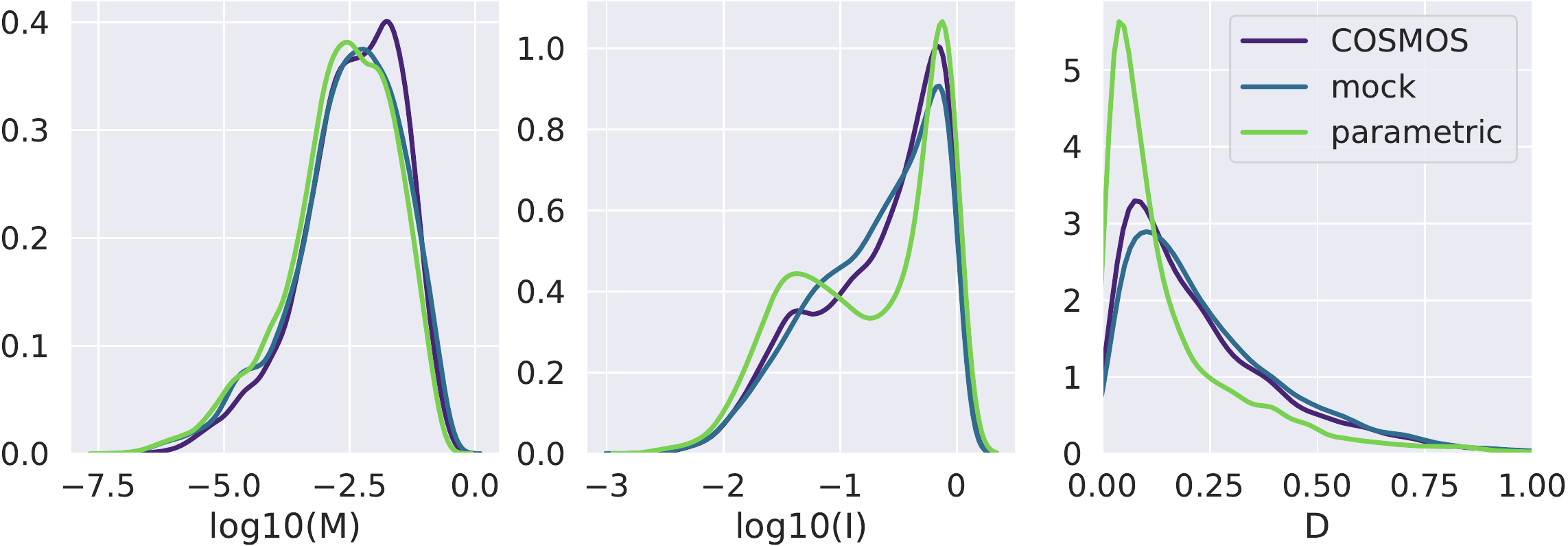} \quad
	\caption{Comparison of marginal MID statistics evaluated on parametric galaxies (left), real COSMOS galaxies (center), samples from the generative model (right). \nblink{Figure_Morphology}}
	\label{fig:MID}
\end{figure}

To further quantitatively compare our generated galaxy sample to the  reference training set, we turn to higher order morphological statistics. In this  work we primarily make use  of the \textit{multi-mode (M)}, \textit{intensity (I)}, and \textit{Deviation (D)} statistics introduced in \cite{Freeman2013}, which are specifically designed to identify disturbed morphologies.  We direct the interested reader to \cite{Freeman2013} for  a thorough description of these statistics and a comparison to standard CAS statistics \citep{Conselice2003}, and we briefly introduce them below:
\begin{itemize}
	\item M(ulti-mode) statistic: detects  multimodality in a galaxy light  profile  as a ratio of area between the largest and second largest contiguous group of pixels above a threshold itself optimized as  to maximize this statistic. M tends to 1 if the light profile exhibits a double nucleus, and  to 0 if the  image is unimodal. 
	\item I(ntensity) statistic: Similar to the M statistic but computes a ratio of integrated flux between  the two most intense contiguous  groups of pixels  in the  image. I tends to 1 for two equally intense nuclei, and to 0 if the flux of the brightest nucleus dominates.
	\item D(eviation) statistic: Measures the distance between the  local intensity maximum identified as  part of the I statistic to the centroid of the light profile computed by a simple first-order moment computation. This distance is scaled by the size of the segmentation map of the object and is  therefore below 1, tending towards 0 for symmetrical galaxies.
\end{itemize}
In addition to these statistics, we also evaluate the  Gini coefficient and M20 statistic \citep{Lotz2004}. These respectively measure the relative distribution of pixel fluxes, and the second-order moment of the brightest 20\% pixels.

\autoref{fig:MID} illustrates the distribution of MID statistics for samples of parametric, mock, and real images. While we do not see a strong deviation in term of the M statistic, the distributions of I and D statistics are significantly different for parametric galaxies, while mock and real galaxies appear to be very similar. 
More specifically, for the I statistic, we note that parametric fits exhibit an under-density around I $ \simeq 0.1$ compared to real COSMOS galaxies. We observed that in this range of I values, multimodal real galaxies are found whereas these do not exist in the monotonic parametric models. As a result, this region is depleted for parametric models. We find that the fits to multimodal COSMOS galaxies from this region are preferentially scattered towards I $=1$ for large structured galaxies, as the modes identified on noisy monotonic profile tend to be from the same neighbourhood and have very similar fluxes. On the other hand, for bright and concentrated galaxies, the parametric fits are scattered to lower I values; in this case the central peak is also clearly identified in the parametric fit, and a second peak, only due to noise, is necessarily artificial and at far lower fluxes. This explains why we observe this bi-modal shape of the $\log($I$)$ distribution of parametric galaxies. 
For the D statistic, we similarly see a significantly higher concentration near D $= 0$ for parametric profiles compared to real COSMOS galaxies.  This is consistent with the definition of this statistic as parametric profiles are symmetric, hence low D statistic. These results for parametric profiles are therefore completely consistent with one's expectations for S\'ersic or Bulge+Disk models with an additional noise field. 

\begin{figure}
    \begin{subfigure}[b]{1.0\columnwidth}
	\includegraphics[width=\columnwidth]{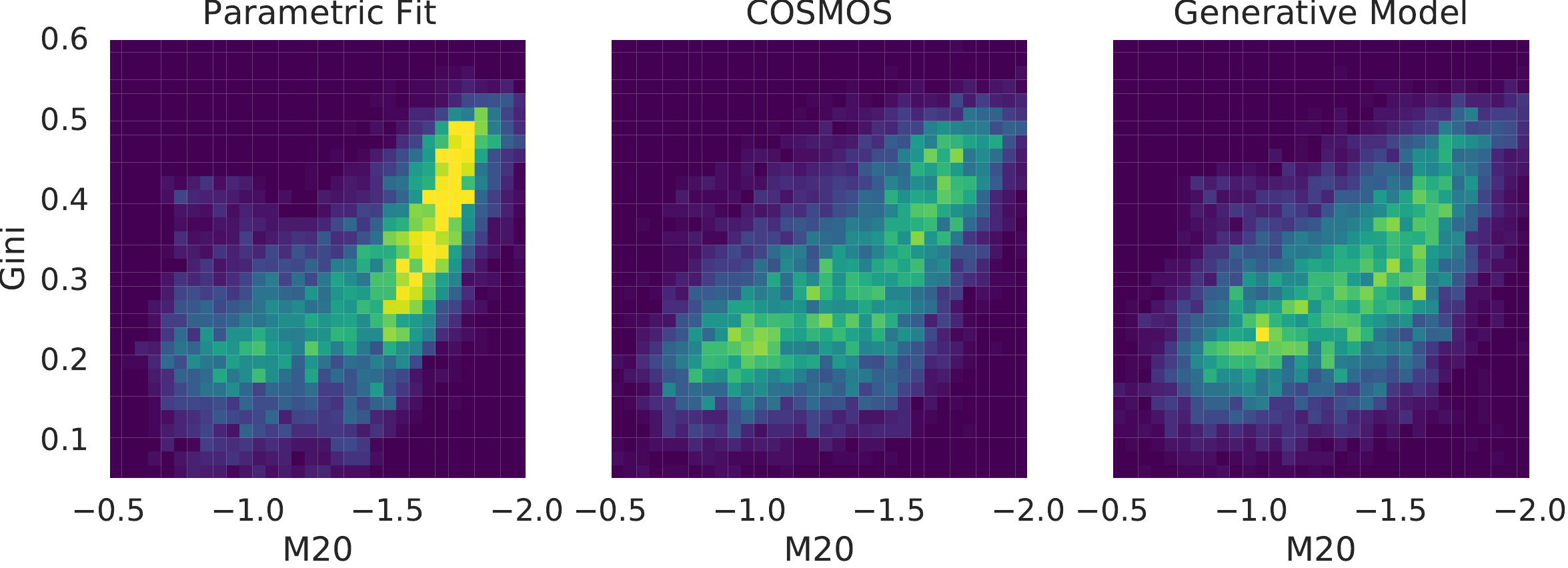}
	\caption{Gini-M20 statistics}
	\label{fig:ginim20a}
	\end{subfigure}\\
	\begin{subfigure}[b]{1.0\columnwidth}
	\includegraphics[width=\columnwidth]{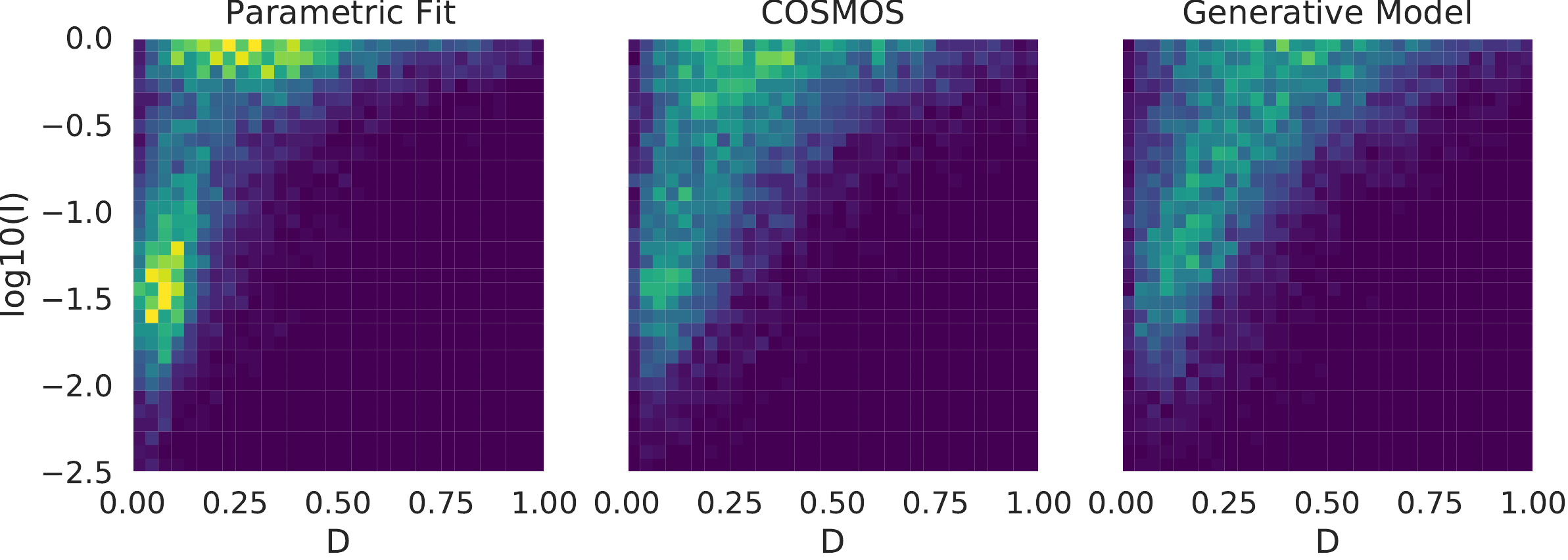}
	\caption{I-D statistics}
	\label{fig:ginim20b}
	\end{subfigure}
	\caption{Comparison of morphology statistics Gini-M20 (a) and I-D (b) evaluated on parametric galaxies (left), real COSMOS galaxies (center), samples from the generative model (right). The colormap is linear. On the Gini-M20 plane, more disturbed morphologies are typically found on the left side of the plot, while smoother morphologies are found to the right.  \nblink{Figure_Morphology}}
	\label{fig:GiniM20}
\end{figure}

By comparison, our mock galaxy images are more consistent with real galaxies, and the fact that they do not exhibit the same failure modes as parametric profiles indicates that the light profiles generated by the deep generative models are indeed less symmetrical and more multimodal than simple profiles. This difference can also be seen in the 2d I-D histograms of \autoref{fig:ginim20b}.

\autoref{fig:ginim20a} provides a similar comparison, but in the Gini-M20 plane, typically used to identify galaxy mergers or galaxies with disturbed morphologies \citep{Lotz2004}. In this plane, galaxies with simpler, less perturbed morphologies are typically found on the right side of the distribution, towards lower M20. On \autoref{fig:ginim20a}, we notice a clear depletion of parametric galaxies at higher M20 and low Gini index (lower left corner) compared to real galaxies. These galaxies seem to have migrated to the right side of the plot, which corresponds to smoother morphologies. We are therefore clearly seeing through this plot that parametric profiles are smoother, less disturbed, than real COSMOS galaxies. On the contrary, no such trend can be identified when comparing COSMOS galaxies to mock galaxies from the Flow-VAE, confirming that under the common Gini-M20 statistic, galaxies sampled from the generative model are also significantly more realistic than simple parametric profiles.

\section{Discussion}
\label{sec:conclusion}

We have presented a framework for building and fitting generative models of galaxy morphology, combining deep learning and physical modeling elements allowing us to explicitly account for the PSF and noise. With this hybrid approach, the intrinsic morphology of galaxies can effectively be decoupled from the observational PSF and noise, which is essential for the use of these generative models in practice. We have further demonstrated a new type of conditional generative model, allowing us to condition galaxy morphology on physical galaxy properties. 
On a sample of galaxies from the HST/ACS COSMOS survey with a limiting magnitude of 25.2 in F814W, we have demonstrated that this deep learning approach to modeling galaxy light profiles not only reproduces distributions of second order moments of the light profiles (i.e. size and ellipticity), but more importantly, is more accurate than conventional parametric light profiles (S\'ersic or Bulge+Disk) when considering a set of morphological summary statistics particularly sensitive to non-monotonicity. We further note that while any deficiencies in modeling second order moments can be trivially addressed by dilation or shearing, these higher order statistics could otherwise not be easily imposed. 

In this section, we now discuss future prospects for applications of these tools as well as further potential improvements and developments.

\bigskip

A first important point highlighted by this work is that when encapsulated within a physical forward model of the instrument, these latent variable generative models can be trained to learn denoised and PSF-deconvolved light profiles. This means that in future work, it will be possible to combine data from ground- and space-based instruments to jointly constrain the same deep and high resolution morphology models. This is to be compared to the current requirement of having access to dedicated deep space based observations which remains limited in quantity and raises concerns such as cosmic variance \citep{Kannawadi2015}. Using HSC deep fields for instance, fitting the morphology model to individual exposures would allow us to profit from the overall depth of the survey as well as from the good seeing exposures bringing more constraints on small scales.

\bigskip

Although we have not emphasized this aspect of our approach in the previous section, the light profiles learned by our models being unconvolved from the PSF, they may contain details beyond the original band-limit of the survey. Thanks to the small amount of Total Variation regularization added to the training loss in \autoref{sec:vae_model}, we find in practice that the model does not introduce obviously unphysical high frequencies or artifacts. Therefore, it \textit{may} be possible to use these galaxy models with a PSF slightly smaller than a typical COSMOS PSF used for training, which can be thought of as some sort of extrapolation to higher band-limits. We caution the user against such a use however, as any details smaller than the original COSMOS resolution are not constrained from data and are purely the results of implicit priors and inductive biases. Testing the impacts of this explicitly, for example by learning a generative model from a version of the COSMOS images degraded in resolution and comparing to the original-resolution images, could be one way to understand the degree to which any extrapolation is possible. This test is left for future work.

As an alternative to learning fully deconvolved light profiles, we also explored partial deconvolution, where galaxies are modeled at a standardized effective PSF only slightly smaller than the training PSFs. In our experiments, although it made the training slightly more stable, it did not significantly affect the performance of the trained model. We did not pursue this option further, but future work using different architectures, especially GANs, may find partial deconvolution advantageous.

\bigskip

Another highlight of this work is the ability to condition galaxy morphology on other physical properties of the object. In an image simulation context, this makes it possible to tie morphology to physical parameters available in mock galaxy catalogs (e.g., stellar mass, colour, magnitude, redshift). This will be crucial for producing complex and realistic survey images accounting jointly for galaxy clustering, photometry, and morphology. 

\bigskip

Beyond image simulations, generative models can be regarded as a general solution for building fully data-driven signal priors which can be used in a range of astronomical imaging inverse  problems such as denoising, deconvolution, or deblending. This idea has been for instance explored in the context of deblending in \cite{Arcelin2020} using a VAE to learn a model of isolated galaxies light profiles, or in \cite{Lanusse2019} using an autoregressive pixelCNN++ \citep{Oord2016} model trained on isolated galaxy images as a prior for deblending by solving a maximum a posteriori optimization problem. The usefulness of latent variable models for solving general inverse problems was further explored in \cite{Boehm2019}, which illustrates how a Flow-VAE such as the one introduced in this work can be used to recover full posteriors on problems such as deconvolution, denoising, and inpainting. 

\bigskip

One open question that has been only partially addressed so far is how to validate the quality of the morphology models. 
As illustrated in this work,  parametric light profiles match by design real galaxies in terms of zeroth, first, and second moments (\autoref{fig:moments}), while metrics based on higher order statistics  (e.g., \autoref{fig:GiniM20}, \autoref{fig:MID}) are able to detect significant departures in morphology. While our particular choice of higher order statistics has proven powerful enough to demonstrate a qualitative gain in morphology over simple parametric profiles,  we have however no guarantee that this set of statistics is sufficient to fully characterize galaxy morphology. Instead of relying on the carefully crafted metrics which are  conventionally used to study galaxy morphologies, recent work has focused on using generative models for anomaly detection. In the first application of these methodologies to astrophysics \citep{Zanisi2020} have for instance demonstrated that a method based on the Log Likelihood Ratio approach of \cite{Ren2019} is capable of identifying morphology discrepancies between IllustrisTNG \citep{Nelson2019} and SDSS  \citep{Abazajian2009, Meert2015} galaxies.

More fundamentally, even if we had access to a set of sufficient statistics to detect deviations between real and generated galaxies, it would remain unclear how close the model would need to match the real morphologies in terms of these statistics in order to satisfy the requirements of a particular scientific application. As an example, let us consider the specific case of calibrating weak lensing shear measurements with image simulations. It is known that the distribution of galaxies ellipticities needs to be modeled with great accuracy \citep{Viola2014}, and precise requirements can be set in terms of ellipticities. These are however necessary but not sufficient conditions; shear couples second-order moments (from which the ellipticity is derived) to higher-order moments of the light profiles \citep{Massey2007c, Bernstein2010, Zhang2011a}, which makes calibration sensitive to morphological details and substructure. Although we have various higher-order statistics at our disposal, defining a set of requirements to ensure accurate calibration is a difficult task and such requirements have never been rigorously quantified in practice.

\bigskip

Finally, here we have proposed a very specific generative model architecture. In our experiments we found this approach of a hybrid VAE and normalizing flow model to be robust and flexible while providing good quality samples. However, we do not expect this model to remain a state-of-the-art solution, and on the contrary we welcome and encourage additional efforts from the community to develop better models. In that spirit, we have put significant efforts into building \texttt{galaxy2galaxy} (\texttt{g2g} for short), a framework for training, evaluating, and exporting generative models on standard datasets such as the COSMOS sample  used in this work. In addition, we have developed the \texttt{galsim\_hub} extension to the \texttt{GalSim} software, which allows us to integrate models trained with \texttt{g2g} directly as GalSim GSObjects which can then be manipulated in the GalSim framework like any other analytic light profile. More details on \texttt{galsim\_hub} can be found in \autoref{sec:galsimhub}.

\bigskip

In the spirit of reproducible and reusable research, the code developed for this paper has been packaged in the form of two Python libraries
\begin{itemize}
	\item Galaxy2Galaxy: Framework for training and exporting generative models
	\begin{center}
		\url{https://github.com/ml4astro/galaxy2galaxy }
	\end{center}
	\item GalSim-Hub: Framework for integrating deep generative models as part of GalSim image simulation software. 
	\begin{center}
		\url{https://github.com/mcwilliamscenter/galsim\_hub }
	\end{center}
\end{itemize}
The scripts used to train the models presented in this work as well as producing all the figures can be found at this link:
\begin{center}
	\url{https://github.com/mcwilliamscenter/deep\_galaxy\_models}
\end{center}

\section*{Acknowledgements}

The authors would like to acknowledge David W. Hogg for useful discussions on hierarchical modeling and comments on a draft of this work, Uro\v{s} Seljak and Vanessa Boehm for countless discussions on generative models, Marc Huertas-Company and Hubert Bretonni\`ere for valuable feedback and testing the framework, and Ann Lee and Ilmun Kim for discussions on statistically comparing galaxy morphologies. FL, RM, and BP were partially supported by NSF grant
IIS-1563887. This work was granted access to the HPC resources of IDRIS under the allocation 2020-101197 made by GENCI. We gratefully acknowledge the support of NVIDIA Corporation with the donation of a Titan Xp GPU used for this research.

\textit{Software: Astropy \citep{Robitaille2013, PriceWhelan2018}, Daft \citep{daft}, GalSim \citep{Rowe2015}, IPython \citep{Perez2007}, Jupyter \citep{Kluyver2016}, Matplotlib \citep{Hunter2007}, Seaborn \citep{seaborn}, TensorFlow \citep{TensorFlow}, TensorFlow Probability \citep{Dillon2017}, Tensor2Tensor \citep{Vaswani2018}}

\section*{Data Availability}

The HST/ACS COSMOS data used in this article are available at \url{http://doi.org/10.5281/zenodo.3242143}. The images and statistics obtained with the generative model are available at \url{http://doi.org/10.5281/zenodo.3975700}.



\bibliographystyle{mnras}
\bibliography{references} 



\appendix
\newpage

\section{GalSim-Hub: online  repository for  trained models}
\label{sec:galsimhub}

As a way to easily interface deep generative models with existing simulation pipelines
based on the \texttt{GalSim} software, we introduce \texttt{GalSim-Hub}: an online repository of
pre-trained models which can directly used within \texttt{GalSim} as any other light profiles.

Concretely, \texttt{GalSim-Hub} is based on the \texttt{TensorFlow Hub}\footnote{\url{https://www.tensorflow.org/hub}} library which allows for TensorFlow models to be saved, loaded, and executed similarly to a conventional Python function within a Python library. In addition to a plain \texttt{TensorFlow Hub} module, \texttt{GalSim-Hub} also specifies some key metadata such as the pixel resolution of the generated image, or input fields required by the module for conditional sampling. At sampling time, the library will generate an \textit{un-convolved} image by drawing from  the generative model, and turn that image into a \texttt{GalSim} \texttt{InterpolatedImage} object which can then be used as any other type of light profile.

To make it easy for researchers to exchange trained deep generative models of galaxy morphology, \texttt{GalSim-Hub} also provides an online repository for community-maintained models directly from the project GitHub repository: \url{https://github.com/mcwilliamscenter/galsim\_hub } .  

\autoref{fig:gshub} illustrates a minimal working example of generating a list of galaxies conditioned on size and magnitude from a pre-trained model available from the online repository.  

\begin{figure}
\begin{Verbatim}[commandchars=\\\{\}]
\PYG{k+kn}{import} \PYG{n+nn}{galsim}
\PYG{k+kn}{import} \PYG{n+nn}{galsim\PYGZus{}hub}
\PYG{k+kn}{from} \PYG{n+nn}{astropy.table} \PYG{k+kn}{import} \PYG{n}{Table}

\PYG{c+c1}{\PYGZsh{} Load generative model from the online repository}
\PYG{n}{model} \PYG{o}{=} \PYG{n}{galsim\PYGZus{}hub}\PYG{o}{.}\PYG{n}{GenerativeGalaxyModel}\PYG{p}{(}
\PYG{l+s+s1}{\PYGZsq{}hub:Lanusse2020\PYGZsq{}}\PYG{p}{)}

\PYG{c+c1}{\PYGZsh{} Defines the input conditions}
\PYG{n}{cat} \PYG{o}{=} \PYG{n}{Table}\PYG{p}{([[}\PYG{l+m+mf}{5.}\PYG{p}{,} \PYG{l+m+mf}{10.} \PYG{p}{,}\PYG{l+m+mf}{20.}\PYG{p}{],}
             \PYG{p}{[}\PYG{l+m+mf}{24.}\PYG{p}{,} \PYG{l+m+mf}{24.}\PYG{p}{,} \PYG{l+m+mf}{24.}\PYG{p}{],}
             \PYG{p}{[}\PYG{l+m+mf}{0.5}\PYG{p}{,} \PYG{l+m+mf}{0.5}\PYG{p}{,} \PYG{l+m+mf}{0.5}\PYG{p}{]],}
             \PYG{n}{names}\PYG{o}{=}\PYG{p}{[}\PYG{l+s+s1}{\PYGZsq{}flux\PYGZus{}radius\PYGZsq{}}\PYG{p}{,} \PYG{l+s+s1}{\PYGZsq{}mag\PYGZus{}auto\PYGZsq{}}\PYG{p}{,} \PYG{l+s+s1}{\PYGZsq{}zphot\PYGZsq{}}\PYG{p}{])}

\PYG{c+c1}{\PYGZsh{} Sample light profiles for these parameters}
\PYG{n}{ims} \PYG{o}{=} \PYG{n}{model}\PYG{o}{.}\PYG{n}{sample}\PYG{p}{(}\PYG{n}{cat}\PYG{p}{)}

\PYG{c+c1}{\PYGZsh{} Define a PSF}
\PYG{n}{psf} \PYG{o}{=} \PYG{n}{galsim}\PYG{o}{.}\PYG{n}{Gaussian}\PYG{p}{(}\PYG{n}{sigma}\PYG{o}{=}\PYG{l+m+mf}{0.06}\PYG{p}{)}

\PYG{c+c1}{\PYGZsh{} Convolve by PSF}
\PYG{n}{ims} \PYG{o}{=} \PYG{p}{[}\PYG{n}{galsim}\PYG{o}{.}\PYG{n}{Convolve}\PYG{p}{(}\PYG{n}{im}\PYG{p}{,} \PYG{n}{psf}\PYG{p}{)} \PYG{k}{for} \PYG{n}{im} \PYG{o+ow}{in} \PYG{n}{ims}\PYG{p}{]}
\end{Verbatim}
\caption{Example of sampling galaxies from the generative model conditioned on size and magnitude with \texttt{GalSim Hub}. The library will automatically download from the online repository models referenced with ``hub:xxxx'' so that no manual user intervention is necessary to run a script.}
\label{fig:gshub}
\end{figure}


\bsp	
\label{lastpage}
\end{document}